\documentclass[aps,preprintnumbers,prd,twocolumn,superscriptaddress,nofootinbib]{revtex4} 

\usepackage{slashed}
\usepackage{graphicx}
\usepackage{color}
\usepackage{amsfonts}
\usepackage{subfigure}
\usepackage{pdflscape}
\usepackage{afterpage}

\newcommand{\bea}{\begin{eqnarray}}
\newcommand{\eea}{\end{eqnarray}}

\newcommand{\ds}{\displaystyle×}

\newcommand{\met}{\,/\hspace{-0.25cm}E_T}

\newcommand{\abinv}{\rm ab$^{-1}$}
\newcommand{\beq}{\begin{equation}}
\newcommand{\eeq}{\end{equation}}

\begin{document}

\preprint{ACFI-T16-12}

\title{Singlet-Catalyzed Electroweak Phase Transitions in the 100~TeV Frontier}
\author{Ashutosh V. Kotwal}
\email{ashutosh.kotwal@duke.edu}
\affiliation{Fermi National Accelerator Laboratory, Batavia, IL 60510, USA}
\affiliation{Department of Physics, Duke University, Durham, NC 27708, USA}
\author{Michael J. Ramsey-Musolf}
\email{mjrm@physics.umass.edu}
\affiliation{Physics Department, University of Massachusetts Amherst, Amherst, MA 01003, USA}
\affiliation{Kellogg Radiation Laboratory, California Institute of Technology, Pasadena, CA 91125, USA}
\author{Jose Miguel No}
\email{J.M.No@sussex.ac.uk}
\affiliation{Department of Physics and Astronomy, University of Sussex, BN1 9QH Brighton, UK}
\author{Peter Winslow}
\email{pwinslow@physics.umass.edu}
\affiliation{Physics Department, University of Massachusetts Amherst, Amherst, MA 01003, USA}

\begin{abstract}
We study the prospects for probing a gauge singlet scalar-driven strong first order electroweak phase transition with a  
future proton-proton collider in the 100~TeV range. Singlet-Higgs mixing enables resonantly-enhanced di-Higgs production, 
potentially aiding discovery prospects. We perform Monte Carlo scans of the parameter space to identify  regions associated with a strong first-order electroweak phase transition, analyze the corresponding di-Higgs signal, and select a set of benchmark points  that span the range of di-Higgs signal strengths. For the $b\bar{b}\gamma\gamma$ and $4\tau$ final states, we investigate discovery prospects  for each benchmark point for the high luminosity phase of the Large Hadron Collider and for a future $pp$ collider with $\sqrt{s}=50$, 100, or 200 TeV. We find that any of these future collider scenarios could significantly extend the reach beyond that of the high luminosity LHC, and that with $\sqrt{s}=100$~TeV (200~TeV) and 30~ab$^{-1}$, the full region of parameter space 
favorable to strong first order electroweak phase transitions is almost fully (fully) discoverable. 
\end{abstract}

\pacs{...}

\maketitle

\section{Introduction}

With the discovery of a Higgs-like boson~\cite{Aad:2012tfa, Chatrchyan:2012ufa} the detailed nature of electroweak symmetry-breaking (EWSB) has come into sharp focus.  While subsequent analyses have shown that the interactions of this new particle closely resemble those expected for the Standard Model 
(SM) Higgs boson, the possibility that it resides within an larger scalar sector remains quite open. Theoretically, an extended scalar sector is motivated by a number of considerations, including  solutions to the hierarchy problem, mechanisms for neutrino mass generation, and dark matter models. 

One of most compelling reasons to postulate an extended scalar sector is to explain the  origin of the baryon asymmetry of the Universe (BAU): 
\begin{eqnarray}
Y_B = \frac{n_B}{s} = (8.59 \pm 0.11) \times 10^{-11} \quad \textrm{(Planck)~\cite{Ade:2013zuv}}
\end{eqnarray}
where $n_B$ ($s$) is the baryon number (entropy) density. It is well known that the SM cannot accommodate the observed BAU, as it fails to provide for both the required CP-violation and the necessary 
 out-of-equilibrium conditions in the early Universe~\cite{Sakharov:1967dj}. While there exist a wide array of scenarios that address these SM shortcomings, one of the most theoretically attractive and experimentally testable is electroweak baryogenesis 
(for a recent review see, e.g.~\cite{Morrissey:2012db}), wherein $Y_B$ is created during the electroweak phase transition (EWPT). Successful electroweak baryogenesis requires that the EWPT be strongly first order.  Monte Carlo lattice simulations indicate that  EWSB in the SM occurs through a cross-over transition~\cite{Aoki:1999fi,Csikor:1998eu,Laine:1998jb,Gurtler:1997hr,Kajantie:1996mn} for a Higgs boson heavier than $70-80$ GeV, thereby precluding electroweak baryogenesis. However, if new bosonic states are present at the electroweak scale, the extra interactions can induce the desired strong first-order  electroweak phase transition (SFOEWPT).

In this study, we investigate the possibility that a next generation proton-proton collider may discover one of the simplest realizations of this possibility: the extension of the SM scalar sector with a single, real gauge singlet,  referred to henceforth as the \lq\lq xSM". As outlined in Ref.~\cite{Profumo:2007wc}, this simple scenario may both accommodate a SFOEWPT and provide for a rich collider phenomenology that may be used to probe it. In general, the xSM yields two mixed doublet-singlet scalars, $h_1$ and $h_2$, that are \lq\lq SM-like"  \lq\lq singlet-like", respectively. 
Among the possible signatures are exotic $h_1$ decays, modifications of the Higgs signal strengths, and resonant production of $h_1$ pairs.  Subsequent work also highlighted the correlation between the SFOEWPT and modifications of the $h_1$ trilinear self-coupling~\cite{Noble:2007kk} and, for a $\mathbb{Z}_2$-symmetric version of the xSM,  production of pairs of singlet scalars that do not mix with the Higgs boson~\cite{Curtin:2014jma}\footnote{Probing the  SFOEWPT in a xSM scenario with an exact $\mathbb{Z}_2$ symmetry is challenging but  may be possible via 
$S$ pair production in vector boson fusion through an off-shell Higgs boson~\cite{Curtin:2014jma}.}. 

Here, we concentrate on resonant di-Higgs production. It is well known that the 
xSM can generate a SFOEWPT in regions of parameter space that parametrically enhance resonant di-Higgs production {\em via} large $h_2 h_1 h_1$ trilinear 
couplings~\cite{Profumo:2007wc,Espinosa:2011ax,No:2013wsa}. Previous work indicates that for relatively light $h_2$, discovery in the $b{\bar b}\tau^+\tau^-$ channel may be possible at the LHC with $\sqrt{s}=14$ TeV and 100 fb$^{-1}$ of integrated luminosity\cite{No:2013wsa}. In this work, we carry out a more comprehensive study, focusing on the
$b{\bar b}\gamma\gamma$ and $4\tau$ states. We find that:
\begin{itemize}
\item A future $pp$ collider with $\sqrt{s}=100$ TeV (200)  could enable discovery of the xSM in nearly all (all) of the SFOEWPT-viable parameter space with 30 ab$^{-1}$ of integrated luminosity.
\item   A future $pp$ collider with $\sqrt{s}=50$ TeV would significantly extend the reach of the high-luminosity phase of the Large Hadron Collider (HL-LHC), but would not provide the comprehensive coverage afforded by a 100 TeV collider.
\item A SFOEWPT could occur in the xSM even if the HL-LHC and a future $e^+e^-$ collider were to constrain the singlet-doublet mixing angle  $| \theta | \lesssim 0.08$.
  In this case, discovery with a 100 TeV $pp$ collider would still remain possible.
\end{itemize}

In arriving at these findings, we first determine the xSM parameter space favorable to a SFOEWPT using Monte Carlo (MC) methods, focusing on the region $m_{2} > 2 \,m_{1}$ where resonant di-Higgs production is kinematically allowed (for an analysis of the region $m_{2} < 2\, m_{1}$, see~\cite{Profumo:2014opa}). 
We then investigate the discovery prospects for resonant di-Higgs production at future $pp$ collider scenarios by identifying a set of 22 benchmark parameter points that span both the SFOEWPT-viable parameter space and the range of associated di-Higgs signal strengths. We focus on the $b\bar{b}\gamma \gamma$ and 4$\tau$ final states, chosen for their clean signatures despite their relatively small cross sections.
We analyze the reach of both the HL-LHC and three  different beam energies for a future $pp$ collider ($\sqrt{s} = 50$, 100, and 200 TeV) as well as several total integrated luminosity goals. 

 We consider detectors with similar performance as the LHC detectors, using this scenario  to set the scale for what 
could be achievable at a next-generation $pp$ collider. We find that both final states studied here provide for comparable sensitivity, with the HL-LHC already being capable of probing 
the larger di-Higgs cross sections in the SFOEWPT parameter space for $m_{2} \lesssim 500$ GeV.  A full exploration of the SFOEWPT-compatible parameter space would require $\sqrt{s}\sim 100$ TeV with 30 fb$^{-1}$. 


Our analysis is organized as follows: in Sec.~\ref{sec:Model} we establish our notation for the xSM and discuss its basic collider 
phenomenology. Section~\ref{sec:bsm} describes the EWPT in the xSM, its related 
phenomenology and our methodology for choosing benchmark points for di-Higgs production. In Sec.~\ref{MCsims}, 
we explore discovery prospects for resonant di-Higgs production for the $b \bar{b} \gamma \gamma$ (Sec.~\ref{bbyy}) and 4$\tau$ (Sec.~\ref{4tau}) 
final states,  and perform a combination of these two channels in Sec.~\ref{Comb}. Finally, in Sec.~\ref{Conclusions}, we present our conclusions.


\section{The xSM: Model and Collider Phenomenology}\label{sec:Model}


In its most general form, the xSM constitutes a framework for simultaneously studying the generic characteristics of singlet scalar driven EWPT dynamics 
and Higgs portal mediated resonant di-Higgs production. The results of this study can thus be mapped onto other models that may also involve additional degrees  of freedom not relevant to either the EWPT or di-Higgs production, {\em e.g.} the NMSSM~\cite{Maniatis:2009re,Ellwanger:2009dp}.
To make the connection between the EWPT dynamics and resonant di-Higgs production, we study the most general form for the xSM zero temperature potential that depends on the Higgs doublet, $H$, and real singlet, $S$ (see {\em e.g.}~\cite{O'Connell:2006wi,Profumo:2007wc,Barger:2007im,Espinosa:2011ax}):
\begin{eqnarray}
\label{ScalarPotential1}
& V(H,S) =\ds  -\mu^2 \left( H^\dagger H \right) + \lambda \left( H^\dagger H \right)^2 + \frac{a_1}{2} \left( H^\dagger H \right) S & \nonumber \\
& \ds + \frac{a_2}{2} \left( H^\dagger H \right) S^2 + \frac{b_2}{2} S^2 + \frac{b_3}{3} S^3 + \frac{b_4}{4} S^4 . &
\label{TotPot}
\end{eqnarray}
The $a_1$ and $a_2$ parameters constitute the Higgs portal that provides the only connection to the SM for the singlet scalar $S$. The $b_2$, $b_3$, 
and $b_4$ parameters are self-interactions that, without the Higgs portal, constitute a hidden sector. In the absence of $a_1$ and $b_3$, the potential has a 
$\mathbb{Z}_2$ symmetry that, if $\left\langle S \right\rangle=0$, stabilizes $S$ and elevates it to the status of a dark matter 
candidate (for a discussion of this possibility, see {\em e.g.}~\cite{Barger:2007im, He:2013suk,Cline:2013gha, Mambrini:2011ik, Gonderinger:2009jp, Espinosa:2008kw, Burgess:2000yq, McDonald:1993ex}).
  However, as both parameters can play a significant role in the strength of the EWPT, they are retained in the  current study, rendering $S$ incapable of acting as a dark matter candidate. For a recent study of the EWPT in the $\mathbb{Z}_2$-symmetric xSM and its signatures at a 100 TeV $pp$ collider, see Ref.~\cite{Curtin:2014jma}.

After EWSB, $H \to (v_0+h) / \sqrt{2}$ with $v_0 = 246$ GeV, and we allow for a possible vacuum expectation value (vev) for $S$, {\em i.e.} $S \to x_0 + s$. Vacuum stability requires the positivity of the quartic coefficients along all directions in field space. Along the $h$ ($s$) direction, this leads to the 
bound $\lambda > 0$ ($b_4 > 0$) while, along an arbitrary direction, this implies $a_2 > - \sqrt{\lambda b_4}$. We note that the sign of any term that breaks the $\mathbb{Z}_2$ symmetry can be changed by the field redefinition $S\to -S$. In the Monte Carlo parameter scan that follows, we will allow the $\mathbb{Z}_2$-breaking operator coefficients $a_1$ and $b_3$ to take on either sign. Doing so is equivalent to fixing the magnitudes of these parameters and carrying out the $S\to -S$ redefinition. Consequently, we will choose $x_0$ to be positive without any loss of generality.

The minimization conditions allow for two of the parameters in Eq.~(\ref{TotPot}) 
to be expressed in terms of the 
vevs and other parameters. For convenience, we choose 
\begin{eqnarray}
& \ds \mu^2 = \lambda v_0^2 + \left( a_1 + a_2 x_0 \right) \frac{x_0}{2} & \nonumber \\
& \ds b_2 = \ds - b_3 x_0 - b_4 x_0^2 - \frac{a_1 v_0^2}{4 x_0} - \frac{a_2 v_0^2}{2} .&
\label{eq:ewsb}
\end{eqnarray}
For viable EWSB, two conditions must be met. The first is that ($v_0,x_0$) is a stable minimum, which requires
\begin{eqnarray}
\ds b_3 x_0 + 2 b_4 x_0^2 - \frac{a_1 v_0^2}{4 x_0} - \frac{ (a_1 + 2 a_2 x_0 )^2 }{ 8 \lambda } > 0 .
\end{eqnarray}
The second is that the EW minimum must be the absolute minimum, which we impose numerically. 

After EWSB, mixing between the states $h$ and $s$ is induced by both the Higgs portal parameters $a_1, a_2$ and the singlet vev with the mass-squared matrix 
\begin{eqnarray}
& m_{h}^2 \equiv \ds \frac{d^2 V}{dh^2} = 2 \lambda v_0^2 & \nonumber \\
& m_{s}^2 \equiv \ds \frac{d^2 V}{ds^2}  = b_3 x_0 + 2 b_4 x_0^2 - \frac{a_1 v_0^2}{4 x_0} & \nonumber \\
& m_{hs}^2 \equiv \ds \frac{d^2 V}{dh ds} = \left(a_1 + 2 a_2 x_0 \right) \frac{v_0}{2} .
\label{mixingM}
\end{eqnarray}
with $m_{hs}^2$ being responsible for the singlet-doublet mixing.
The corresponding mass eigenstates are given by
\begin{eqnarray}
& h_1 = h \cos \theta + s \sin \theta & \nonumber \\
& h_2 = - h \sin \theta + s \cos \theta &
\label{eigstates}
\end{eqnarray}
where $h_1$ ($h_2$) is the more $SU(2)_L$-like (singlet-like) scalar and the mixing angle $\theta$ is most easily defined in terms of the mass eigenvalues,
\begin{eqnarray}
& m_{2,1}^2 = \ds \frac{ m_{h}^2 + m_{s}^2 \pm \left| m_{h}^2 - m_{s}^2 \right| \sqrt{ 1 + \ds \left( \frac{ m_{hs}^2 }{ m_{h}^2 - m_{s}^2 } \right)^2 } } {2} ,& \nonumber \\
\label{Meigenvalues}
\end{eqnarray}
as
\begin{eqnarray}
\sin 2 \theta =  \ds \frac{ 2 m_{hs}^2 }{ \left( m_1^2 - m_2^2 \right) } = \ds \frac{ \left( a_1 + 2 a_2 x_0 \right) v_0  }{ \left( m_1^2 - m_2^2 \right) } .
\label{sin2theta}
\end{eqnarray}
The $SU(2)_L$-like scalar eigenstate $h_1$ is considered the lighter eigenstate and identified with the observed Higgs boson at the LHC~\cite{Aad:2012tfa, Chatrchyan:2012ufa}, 
so we set $m_1 = 125$ GeV. According to Eq.~(\ref{eigstates}), the couplings of $h_1$ and $h_2$ to all SM states are simply rescaled versions of SM Higgs couplings, i.e., 
\begin{eqnarray}
g_{h_1 XX} = \cos \theta \; g_{h XX}^{\mathrm{SM}}, \quad g_{h_2 XX} = \sin \theta \; g_{h XX}^{\mathrm{SM}}
\end{eqnarray}
with $XX$ representing any SM final state.
The mixing angle is thus constrained by measurements of Higgs signal strengths, oblique parameters, and direct heavy SM-like Higgs searches. 
 A SFOEWPT requires  $m_{hs}^2 < 0 $ and correspondingly $\sin 2 \theta >0$. 
 
\begin{figure*}[t!]
\centering
\mbox{
\subfigure{
\hspace{-.25in}\includegraphics[scale=.65]{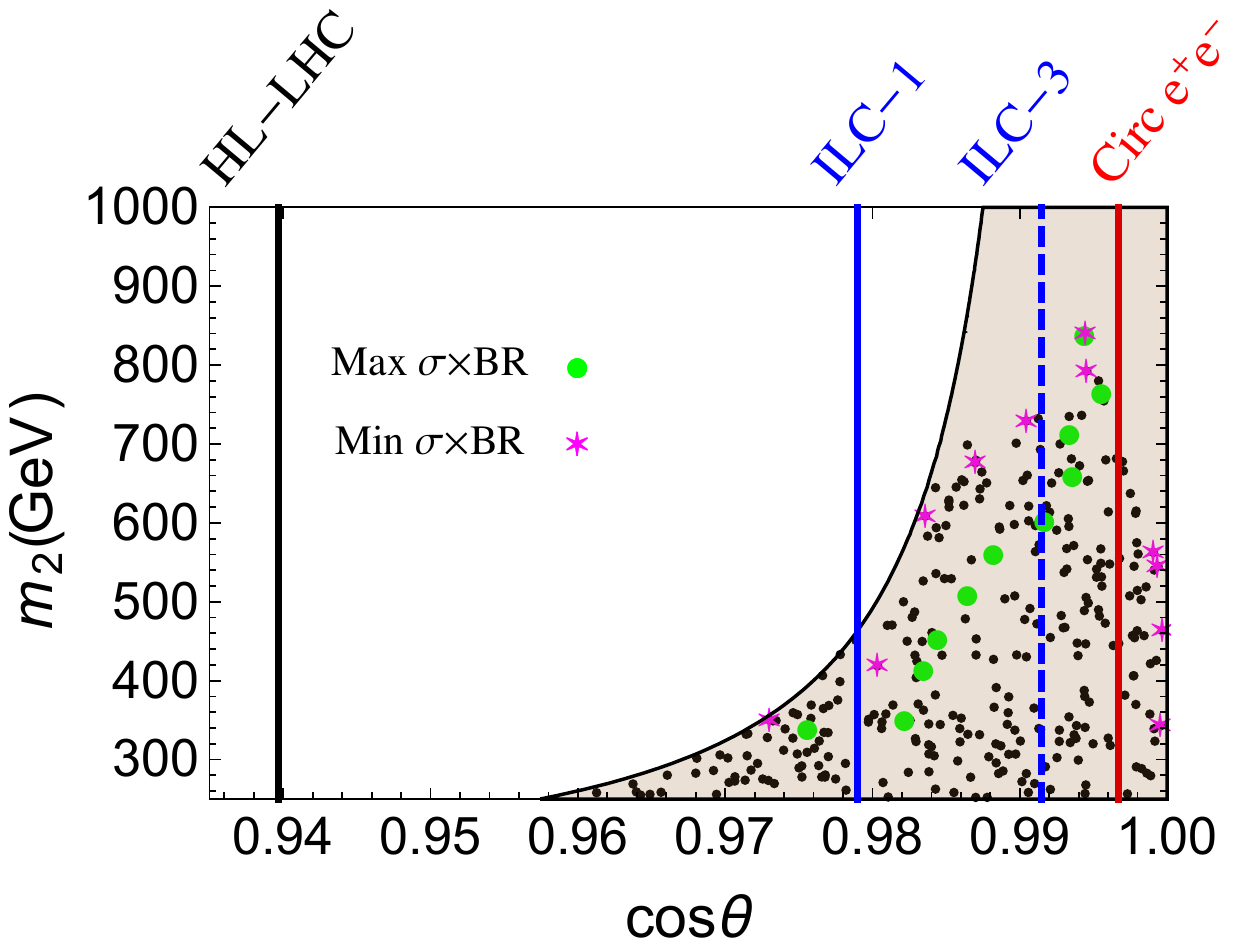}
}
\subfigure{
\hspace{.25in}\includegraphics[scale=.6]{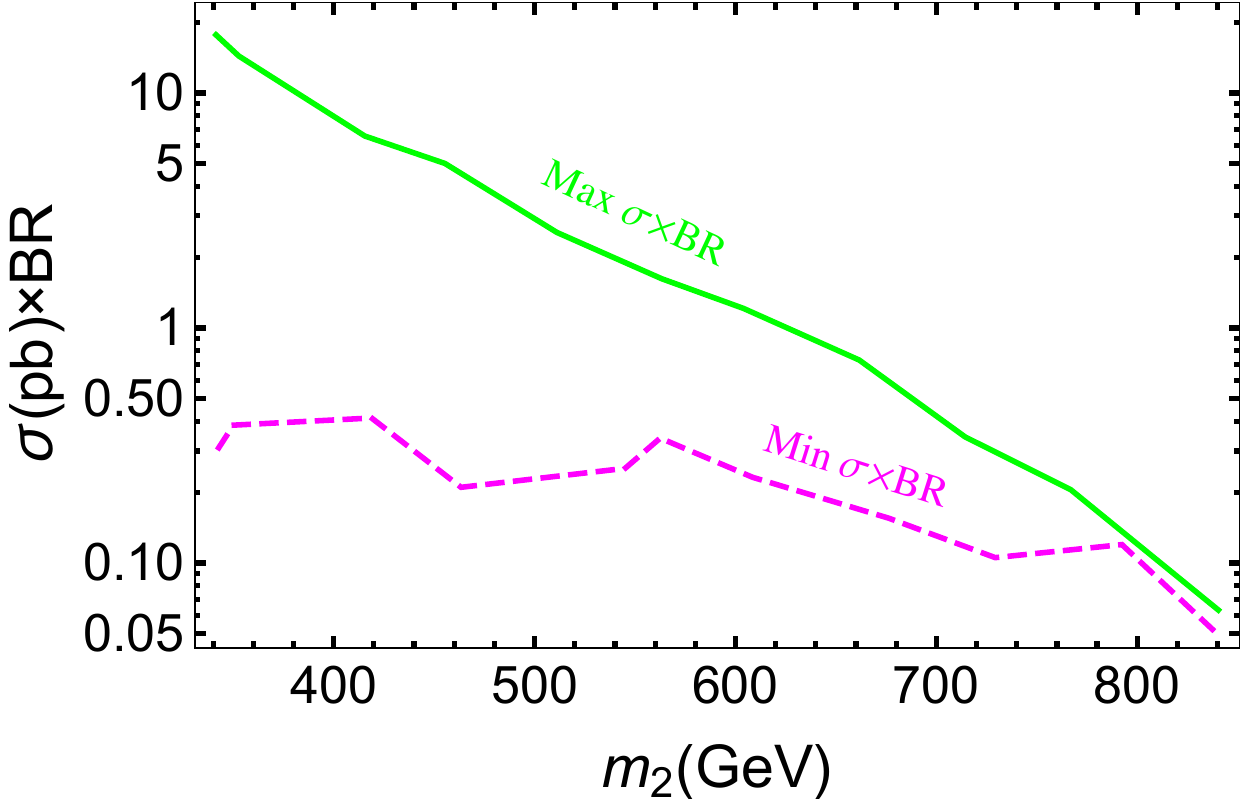}
}
}
\caption{\small Left pane: Distribution of SFOEWPT points in $m_2$ {\it vs} $\cos \theta$ space. Maximum (minimum) benchmark points are shown in green (magenta). Right pane: Maximum (minimum) cross section times branching ratio as a function of $m_2$ at a 100~TeV $pp$ collider, taken from Table~\ref{T1} (Table~\ref{T2}), is displayed as a solid green (dashed magenta) line. }
\label{m2_VS_cos}
\end{figure*}

In this work, we 
concentrate on the kinematic regime in which resonant di-Higgs production occurs, which we take to be $2\, m_1 \leq m_2 \leq$ 1 TeV. 
In this case, $h_1$ has no new scalar decay modes, which implies that all signal rates associated with Higgs measurements are functions of the mixing angle only:
\begin{eqnarray}
\mu_{h_1 \to XX} = \frac{\sigma \cdot \text{BR}}{\sigma^{SM} \cdot \text{BR}^{\mathrm{SM}}} = \cos^2 \theta
\end{eqnarray}
where $\sigma$ is the production cross section and BR is the branching ratio (equal to BR$^{\mathrm{SM}}$ in the absence of new $h_1$ scalar decay modes). 
The current limit from Higgs measurements, obtained by performing a global $\chi^2$ fit to data from both ATLAS and CMS, is 
$|\cos \theta| \gtrsim$ 0.85~\cite{Profumo:2014opa}. 
Estimated sensitivities to the mixing angle from future collider experiments may also be obtained 
using a simple $\chi^2$-method (see~\cite{Profumo:2014opa,Gorbahn:2015gxa} for details). As in~\cite{Profumo:2014opa}, we 
derive projected sensitivities for the high luminosity LHC ($\sqrt{s}$ = 14 TeV, 3~ab$^{-1}$), the ILC 
(ILC-1: $\sqrt{s}$ = 250 GeV, 250~fb$^{-1}$ and ILC-3: $\sqrt{s}$ = 1 TeV, 1~ab$^{-1}$), and 
 a future circular $e^+e^-$ collider  ($\sqrt{s}$ = 240 GeV, 1~ab$^{-1}$), shown in Fig.~\ref{m2_VS_cos} (left) as black, blue, and red vertical lines respectively.

The effects of the xSM on electroweak precision observables and the $W$-boson mass are characterized by the oblique parameters $S$, $T$, and $U$. 
From Eq.~(\ref{eigstates}), the shift in any oblique parameter, $\mathcal{O}$, can be written entirely in terms of the SM Higgs contribution to that 
parameter, $\mathcal{O}^{SM}(m)$, where $m$ is either $m_1$ or $m_2$. These shifts then take the form
\begin{eqnarray}
& \Delta \mathcal{O} = (\cos^2 \theta -1) \mathcal{O}^{SM}(m_1) + \sin^2 \theta \; \mathcal{O}^{SM}(m_2) & \nonumber \\
& = \sin^2 \theta \left( \mathcal{O}^{SM}(m_2) - \mathcal{O}^{SM}(m_1) \right) ,&
\end{eqnarray}
where it is clear that the corresponding constraint is significantly enhanced in the high mass region. We take the best fit values for the shifts, 
$\Delta \mathcal{O}$, from the most recent post-Higgs-discovery electroweak fit to the SM by the Gfitter group~\cite{Baak:2012kk} and perform a global 
$\chi^2$ fit, including all correlations, to this data (for details, see~\cite{Profumo:2014opa}). The 95\% C.L. allowed region in the ($\cos \theta, m_2$) 
plane is shown in Fig.~\ref{m2_VS_cos} (left) as the beige shaded region.

LHC searches for a heavy SM-like Higgs boson also provide a probe of $h_2$ since it will decay to all SM Higgs boson decay products as well as to $h_1$ pairs (for $m_2> 2m_1$).  In particular, the 
 ATLAS~\cite{Aad:2015kna,Aad:2015ipg} and CMS~\cite{Chatrchyan:2013yoa,Khachatryan:2015cwa} Collaborations
  have  performed searches for SM-like heavy Higgs bosons in the mass range 145-1000 GeV focusing on $WW$ and $ZZ$ final  states, placing limits on the corresponding signal rate at the 95\% C.L. 

All production modes for $h_2$ are inherited entirely from mixing and, thus, $\sin \theta$ fully controls the production cross section with respect to
(w.r.t.) its SM value. 
In contrast, in the kinematic regime where resonant di-Higgs production is allowed, the new scalar decay mode $h_2 \to h_1 h_1$ yields a modification 
of all the $h_2$ branching fractions w.r.t. their SM values. This new decay mode is dependent on the trilinear coupling
\bea
\label{g211}
\lambda_{211} = \frac{1}{4}\left[ (a_1+2a_2x_0) \cos^3 \theta + 4 v_0 (a_2 -3 \lambda) \cos^2 \theta \sin \theta \right. \nonumber\\
\left. + (a_1+2a_2x_0 -2b_3-6b_4x_0) \cos \theta \sin^2 \theta -2a_2 v_0 \sin^3 \theta \right] \nonumber \\
\eea
and, along with the $\sin^2\theta$ rescaling, modifies the rate associated with the heavy Higgs production and decay. The partial width  $\Gamma_{h_2 \to h_1 h_1}$ 
is given by
\begin{eqnarray}
\Gamma_{h_2 \to h_1 h_1} = \frac{\lambda_{211}^2 \sqrt{1 - \ds 4 m_1^2 / m_2^2 } }{8 \pi m_2} . 
\label{partialWidthh1h1}
\end{eqnarray}
Defining $\Gamma^{\mathrm{SM}}(m_2)$ as the SM Higgs width evaluated at $m_2$, which we take from~\cite{Heinemeyer:2013tqa},  the total width for the $h_2$ boson is given by 
\begin{eqnarray}
\Gamma_{h_2} = \mathrm{sin}^2 \theta \; \Gamma^{\mathrm{SM}}(m_2) + \Gamma_{h_2 \to h_1 h_1} .
\end{eqnarray}
The resulting signal rate  (normalized to the SM value) 
for $p p \to h_2 \to X X$  (with $XX$ representing all SM final states except $h_1 h_1$) is
\begin{eqnarray}
\mu_{h_2 \to XX} = \sin^2 \theta \left( \frac { \sin^2 \theta \; \Gamma^{\mathrm{SM}}(m_2)}{ \Gamma_{h_2} }  \right).
\end{eqnarray}
 Due to the implicit dependence on $\lambda_{211}$, it is not possible to display the CMS constraint on $\mu_{h_2 \to XX}$ in the form of a smooth region in Fig.~\ref{m2_VS_cos} (left). 
However, we apply this constraint at the level of a MC scan (see Sec. \ref{sec:bsm}) and find that doing so excludes no additional parameter regions beyond those already ruled out by electroweak precision observables at the 95\% C.L.


\section{Electroweak Phase Transition \& Benchmarks 
for di-Higgs Production}
\label{sec:bsm}

The character of the EWPT is understood in terms of the behavior of the finite temperature effective potential, $V_{\rm eff}^{T\neq 0}$. However, it is well-known that 
the standard derivation of $V_{\rm eff}^{T\neq 0}$ suffers from gauge dependence (see~\cite{Patel:2011th} for an in-depth review). Although the value of the EWSB 
vev at the critical temperature, $\phi (T_c)$, is inherently gauge-dependent as it is not an observable, the standard method for extracting $T_c$ also introduces a 
separate and spurious gauge-dependence. The consequence is that the conventional criterion for avoiding baryon washout, $\phi(T_c) / T_c \gtrsim 1$, inherits both 
sources. In this work, we employ a high temperature expansion to restore gauge-indepence to our analysis (see~\cite{Profumo:2014opa} for details). This 
requires forgoing the addition of the $T$=0 Coleman-Weinberg 1-loop effective potential as well as retaining only the gauge-independent thermal mass corrections 
to $V_{\rm eff}^{T\neq 0}$, which are critical to restoring electroweak symmetry at high temperatures. Within this limit, the $T$-dependent vevs, $T_c$, and the 
bubble nucleation rate are all manifestly gauge-independent. We note here that this limit is particularly suited to the xSM, which generates the required barrier 
between the broken and unbroken phases at tree-level via the parameters $a_1$ and $b_3$.

In the high temperature limit, we follow~\cite{Pietroni:1992in} and write the $T$-dependent, gauge-independent (indicated by the presence of a bar) vevs in a 
cylindrical coordinate representation as
\begin{eqnarray}
\bar{v}(T) / \sqrt{2} = \bar{\phi} \cos \alpha (T), \quad \bar{x}(T) = \bar{\phi} \sin \alpha (T)\, .
\end{eqnarray}
The energy of the electroweak sphaleron responsible for baryon washout is proportional to the SU(2)$_L$-breaking energy scale, $\bar{v}(T)$. Sufficient 
quenching of the sphaleron transitions in the broken phase, in order to preserve any baryon asymmetry against washout, is then characterized by the requirement
\begin{eqnarray}
\label{eq:vctc}
\cos \alpha (T_c ) \,\frac{\bar{\phi} (T_c)}{T_c} \gtrsim 1 \,.
\end{eqnarray}
If this condition is met, then the EWPT is said to be a SFOEWPT. As emphasized in Ref.~\cite{Patel:2011th}, this criterion is subject to a number of theoretical uncertainties, even in the presence of a gauge-invariant computation as performed here. Consequently, when considering the phenomenological implications resulting from our parameter scan, one should treat constraints imposed by Eq.~(\ref{eq:vctc}) as approximate.

The critical values, $\bar{\phi}(T_c)$ and $\alpha(T_c)$, are determined by minimizing 
$V_{\rm eff}^{T\neq 0} (\phi, \alpha, T)$ while $T_c$ is defined as the temperature at which the broken and unbroken phases are degenerate:
$V_{\rm eff}^{T \neq 0} (\phi, \alpha \neq \pi/2, T_c) = V_{\rm eff}^{T \neq 0} (\phi, \alpha = \pi/2, T_c)$. We implement the xSM in the high temperature 
limit in {\sc CosmoTransitions}~\cite{Wainwright:2011kj} to numerically obtain all of the above quantities characterizing the EWPT. Moreover, we 
calculate the finite temperature thermal tunnelling rate into the electroweak phase, requiring it to be sufficiently fast in order to preclude the 
possibility of the Universe becoming stuck in a false metastable phase. 

With these considerations, we take $a_1$, $b_3$, $x_0$, and $b_4$ as independent parameters and perform MC scans of the xSM parameter space within the following ranges
\begin{eqnarray}
& a_1/\text{TeV}, b_3/\text{TeV} \in [-1, 1], \quad x_0/\text{TeV} \in [0, 1], & \nonumber \\
& b_4, \lambda \in [0, 1] &
\end{eqnarray}
where the lower bounds on the quartic couplings ensure vacuum stability. With our choice of independent 
 parameters, both $\cos \theta$ and $a_2$ are  fixed by the parameters of the scan and $m_2$. Following Ref.~\cite{Curtin:2014jma}, we impose a na\"ive perturbativity bound
on the Higgs portal coupling $a_2/2 \lesssim 5$.
For each point, we require a SFOEWPT with sufficient tunneling rate as well as 
consistency with all basic theoretical limits and current bounds from Higgs measurements, electroweak precision observables, and heavy Higgs searches. 
We present all points that pass these requirements (displayed in black) in the ($\cos \theta$, $m_2$) plane in Fig.~\ref{m2_VS_cos} (left). 
 Further details about the generic parameteric behavior under the above conditions are provided in~\cite{Profumo:2014opa}.


We turn now to the analysis of resonant di-Higgs production, beginning with our selection of benchmark points. We focus on gluon fusion, as it is by far the dominant production 
mechanism within the $m_2$ range of interest. As already stressed above, xSM di-Higgs production differs considerably from the analogous process in the SM, 
 since the $s$-channel 
$gg \to h_2 \to h_1 h_1$ amplitude is resonant for $m_2 > 2\, m_1$, leading to a large enhancement of the production cross 
section as well as a different kinematic behavior of the full di-Higgs amplitude. 
In addition, di-Higgs production in the SM and xSM differ in two important ways: 
{\it (i)} For the xSM, the 1-loop $gg\, h_2$ interaction is rescaled by $\sin \theta$, leading to a suppression of the cross section by $\sin^2 \theta$. 
{\it (ii)} The trilinear coupling involved in producing the $h_1 h_1$ final state is different depending on whether $h_1$ or $h_2$ is the intermediate state. 
Moreover, the $h_1^3$ trilinear coupling $\lambda_{111}$ in the xSM can also differ significantly from its SM 
value within the parameter space leading to a SFOEWPT~\cite{Profumo:2014opa}. 

For $\lambda_{211} = 0$ the branching fractions of $h_2$ into SM states equal those of a SM Higgs boson with mass $m_{2}$ 
(recall the discussion at the end of Section~\ref{sec:Model}). For $\lambda_{211} \neq 0$,  the branching fraction for $h_2 \to h_1 h_1$ 
\beq
\mathrm{BR}(h_2 \to h_1 h_1) = \frac {\Gamma_{h_2 \to h_1 h_1}} {\Gamma_{h_2}}  
\label{Br211}
\eeq
incorporates a  non-trivial parameter dependence through $\lambda_{211}$ since 
  the partial width  $\Gamma_{h_2 \to h_1 h_1}$ is proportional to $ \lambda_{211}^2 $ (see Eqn.~\ref{partialWidthh1h1}). 

The resonant di-Higgs cross section is thus given at leading order (LO) 
by~$\sin^2 \theta \times \sigma_{\mathrm{LO}} (p p \to h )^{\mathrm{SM}} (m_2) \times \mathrm{BR} (h_2 \to h_1 h_1)$. 
Following~\cite{Djouadi:2005gi}, we write $\sigma_{\mathrm{LO}} (p p \to h )^{\mathrm{SM}} (m_2)$ as
\begin{eqnarray}
& \sigma_{\mathrm{LO}} (p p \to h )^{\mathrm{SM}}(m_2) = \ds \frac{G_F \,\alpha_s^2}{512 \sqrt{2} \pi} \left| \sum_q A_{1/2} \left( \frac{m_2^2}{4 m_q^2} \right) \right|^2  
& \nonumber \\
& \times \; m_2^2 \,\ds \frac{d \mathcal{L}}{d m_2^2} &
\label{LOsigma}
\end{eqnarray}
where $G_F$ is the Fermi constant, $\alpha_s$ is the strong coupling (evaluated at 100 TeV), and $A_{1/2}$ is the loop function  given in~\cite{Djouadi:2005gi}. 
In the case of resonant production, the convolution of parton distribution functions with the LO cross section yields a single parton luminosity function 
$\frac{d \mathcal{L}}{d m_2^2}$ (given {\em e.g.} in~\cite{Heinemeyer:2013tqa}) for energies $\sqrt{s}$ = 100 TeV and Higgs mass values of throughout 
the $m_2$ range of interest. 
Our results at LO in QCD are expected to be conservative estimates of signal sensitivity, as higher-order contributions, encoded in the relevant $k$-factors, would increase both signal and background cross sections
 and increase the sensitivity by $\sim\sqrt{k}$. 

Using the results in Eqs.~(\ref{Br211}) and (\ref{LOsigma}), we choose two sets of benchmark points from our previous MC scan of the xSM parameter space. The 
first set, labeled BM$^\mathrm{max}$, consists of the points that maximize the LO di-Higgs rate in each 50 GeV window within the range $m_2 \in$ [300 GeV, 1 TeV]. 
The second set, labeled BM$^\mathrm{min}$, is analogous to the first but for points that minimize the LO di-Higgs rate. We show both sets in Fig.~\ref{m2_VS_cos} (left), 
with BM$^\mathrm{max}$ as green circles and BM$^\mathrm{min}$ as magenta stars, and display their numerical values respectively in Tables~\ref{T1} and~\ref{T2}. Also shown in 
 Fig.~\ref{m2_VS_cos} (right) are the maximum and minimum cross section times branching ratio as a function of $m_2$, corresponding to these benchmark points. To guide the reader's eye and indicate the overall trends, we have connected the BM$^\mathrm{max}$ (BM$^\mathrm{min}$) di-Higgs cross sections with a solid green (dashed magenta) line.

It is worth stressing that it is possible to find highly-tuned points in the xSM parameter space that yield a SFOEWPT while 
featuring a very fine cancellation among different terms in~(\ref{g211}), 
leading to $\lambda_{211} \to 0$. Such ``outlier" points would thus yield a value for $\sigma\times$BR much below a sensible BM$^\mathrm{min}$, 
but they correspond to very tuned corners of the xSM that do not represent the general properties of the model. In our MC scan, these outliers can be identified as yielding a dramatic 
 drop in $\sigma \times$BR with respect to the subsequent BM$^\mathrm{min}$ candidate benchmark within each 50 GeV mass window. We have identified and eliminated one such outlier 
 point in favor of the selected BM8$^\mathrm{min}$.

\begin{table*}[t!]
  \begin{tabular}{| c | c | c | c | c | c | c | c | c| c| c| c | c| c| c | c |}
    \hline
 Benchmark & ~ $\cos \theta$ ~ & ~ $\sin \theta$ ~ & ~ $m_2$ ~ & ~ $\Gamma_{h_2}$ ~  & ~ $x_0$ ~ & ~ $\lambda$ ~ & ~ $a_1$ ~ & ~ $a_2$ ~ & ~ $b_3$ ~ & ~ $b_4$ ~ & ~ $\lambda_{111}$ ~ & ~ $\lambda_{211}$ ~ & ~ $\sigma$ ~  & ~ BR ~ \\         
           &               &               & (GeV) &     (GeV)        & (GeV) &           & (GeV) &       & (GeV) &       &      (GeV       &       (GeV)     &  ~  (pb) ~  &  \\         
    \hline
B1 & 0.976 & 0.220 & 341 & 2.42 &  257 & 0.92 & -377 & 0.392 & -403 & 0.77 & 204 & -150 & 23.9 & 0.74 \\
B2 & 0.982 & 0.188 & 353 & 2.17 &  265 & 0.99 & -400 & 0.446 & -378 & 0.69 & 226 & -144 & 19.0 & 0.76 \\
B3 & 0.983 & 0.181 & 415 & 1.59 &  54.6 & 0.17 & -642 & 3.80 & -214 & 0.16 & 44.9 & 82.5 & 20.1 & 0.33 \\
B4 & 0.984 & 0.176 & 455 & 2.08 &  47.4 & 0.18 & -707 & 4.63 & -607 & 0.85 & 46.7 & 93.5 & 16.3 & 0.31 \\
B5 & 0.986 & 0.164 & 511 & 2.44 &  40.7 & 0.18 & -744 & 5.17 & -618 & 0.82 & 46.6 & 91.9 & 10.8 & 0.24 \\
B6 & 0.988 & 0.153 & 563 & 2.92 &  40.5 & 0.19 & -844 & 5.85 & -151 & 0.083 & 47.1 & 104 & 6.96 & 0.23 \\
B7 & 0.992 & 0.129 & 604 & 2.82 &  36.4 & 0.18 & -898 & 7.36 & -424 & 0.28 & 45.6 & 119 & 4.01 & 0.30 \\
B8 & 0.994 & 0.113 & 662 & 2.97 &  32.9 & 0.17 & -976 & 8.98 & -542 & 0.53 & 44.9 & 132 & 2.23 & 0.33 \\
B9 & 0.993 & 0.115 & 714 & 3.27 &  29.2 & 0.18 & -941 & 8.28 & 497 & 0.38 & 44.7 & 112 & 1.73 & 0.20 \\
B10 & 0.996 & 0.094 & 767 & 2.83 &  24.5 & 0.17 & -920 & 9.87 & 575 & 0.41 & 42.2 & 114 & 0.918 & 0.22 \\
B11 & 0.994 & 0.105 & 840 & 4.03 & 21.7 & 0.19 & -988 & 9.22 & 356 & 0.83 & 43.9 & 83.8 & 0.802 & 0.079 \\
 \hline
  \end{tabular}
  \caption{\small Values of the various xSM independent and dependent parameters for each of the benchmark values chosen to {\bf maximize} the $\sigma \cdot BR (h_2 \to h_1 h_1)$ at a 100~TeV $pp$ collider. 
  These values are represented as green circular points in Fig.~\ref{m2_VS_cos} (left) and as the solid green curve in Fig.~\ref{m2_VS_cos} (right).}
\label{T1}
\vspace{6mm}
\end{table*} 

\begin{table*}[t!]
  \begin{tabular}{| c | c | c | c | c | c | c | c | c| c| c| c | c| c| c | c |}
    \hline
 Benchmark & ~ $\cos \theta$ ~ & ~ $\sin \theta$ ~ & ~ $m_2$ ~ & ~ $\Gamma_{h_2}$ ~  & ~ $x_0$ ~ & ~ $\lambda$ ~ & ~ $a_1$ ~ & ~ $a_2$ ~ & ~ $b_3$ ~ & ~ $b_4$ ~ & ~ $\lambda_{111}$ ~ & ~ $\lambda_{211}$ ~ & ~ $\sigma$ ~  & ~ BR ~ \\         
           &               &               & (GeV) &     (GeV)        & (GeV) &           & (GeV) &       & (GeV) &       &      (GeV       &       (GeV)     &  ~  (pb) ~  &  \\         
    \hline
B1 & 0.999 & 0.029 & 343 & 0.041 &  105 & 0.13 & -850 & 3.91 & -106 & 0.29 & 32.1 & 19.3 & 0.428 & 0.72 \\
B2 & 0.973 & 0.231 & 350 & 0.777 &  225 & 0.18 & -639 & 0.986 & -111 & 0.97 & 37.7 & 11.6 & 27.8 & 0.014 \\
B3 & 0.980 & 0.197 & 419 & 1.32 &  234 & 0.18 & -981 & 1.56 & 0.42 & 0.96 & 39.0 & 17.5 & 23.5 & 0.018 \\
B4 & 0.999 & 0.026 & 463 & 0.0864 &  56.8 & 0.13 & -763 & 6.35 & 113 & 0.73 & 32.2 & 27.4 & 0.334 & 0.63 \\
B5 & 0.999 & 0.035 & 545 & 0.278 &  50.2 & 0.13 & -949 & 8.64 & 151 & 0.57 & 33.0 & 51.6 & 0.408 & 0.62 \\
B6 & 0.999 & 0.043 & 563 & 0.459 &  33.0 & 0.13 & -716 & 9.25 & -448 & 0.96 & 33.7 & 66.8 & 0.553 & 0.62 \\
B7 & 0.984 & 0.180 & 609 & 4.03 &  34.2 & 0.22 & -822 & 4.53 & -183 & 0.57 & 47.8 & 45.2 & 7.67 & 0.030 \\
B8 & 0.987 & 0.161 & 676 & 4.47 &  30.3 & 0.22 & -931 & 5.96 & -680 & 0.43 & 48.4 & 55.2 & 4.17 & 0.037 \\
B9 & 0.990 & 0.138 & 729 & 4.22 &  27.3 & 0.21 & -909 & 6.15 & 603 & 0.93 & 45.7 & 61.0 & 2.33 & 0.045 \\
B10 & 0.995 & 0.104 & 792 & 3.36 &  22.2 & 0.18 & -936 & 9.47 & -848 & 0.66 & 43.5 & 92.4 & 0.991 & 0.12 \\
B11 & 0.994 & 0.105 & 841 & 3.95 &  21.2 & 0.19 & -955 & 8.69 & 684 & 0.53 & 43.3 & 73.4 & 0.801 & 0.062 \\
 \hline
  \end{tabular}
  \caption{\small Values of the various xSM independent and dependent parameters for each of the benchmark values chosen to {\bf minimize} the $\sigma \cdot BR (h_2 \to h_1 h_1)$ at a 100~TeV $pp$ collider. 
  These values are represented as magenta star-shaped points in Fig.~\ref{m2_VS_cos} (left) and as the dashed magenta curve in Fig.~\ref{m2_VS_cos} (right).}
\label{T2}
\end{table*}

We note here that no SFOEWPT-viable points are discovered by the scan above $m_2 \sim 850$ GeV even though it accepts points up to $m_2 =$ 1 TeV. Moreover, it is clear from Fig.~\ref{m2_VS_cos} (left)
that (i)  prospective circular $e^+e^-$ colliders are expected to have the ability to probe all benchmark points in  BM$^\mathrm{max}$; (ii) the ILC could probe up to  BM8$^\mathrm{max}$; and (iii) neither $e^+e^-$ collider option 
has the capability of excluding the full SFOEWPT-viable xSM parameter space. In short, many points in BM$^\mathrm{min}$ lie beyond 
the sensitivity reach of presently envisioned, future $e^+e^-$ colliders. In the next section, we show that there are options for future $pp$ colliders  in the 100 TeV range that would be capable of discovering 
not only the benchmarks of BM$^\mathrm{max}$ but also those of BM$^\mathrm{min}$, rendering the entire SFOEWPT-viable xSM parameter space discoverable.

\section{Monte Carlo simulations}
\label{MCsims}

The prospective final states  for searching for  di-Higgs production are reproduced from~\cite{Kotwal:2015rba} in Table~\ref{branchingRatios}, ranked by branching ratio. 
In this table, the gauge bosons are required to decay to leptons to suppress enormous backgrounds from QCD jet production. 
QCD production of $b$-jets and $t \bar{t}$ respectively will likely overwhelm the $4b$ channel, as well as 
make the $b\bar{b}\tau\tau$ and $b\bar{b} W^{+}W^{-}$ ($W \to \ell \nu$) channels challenging. 
Motivated by these considerations, we study the $b\bar{b}\gamma\gamma$ and $4\tau$ final states, which are complementary 
having different backgrounds and mass resolution, and discuss the combined results in Sec.~\ref{Comb}.

After applying kinematic and fiducial cuts, we use the expected signal and background event rates and distributions to estimate the discovery potential of a future 100-TeV scale 
$pp$ collider. We use the {\sc Madgraph5}~\cite{Alwall:2011uj}  event generator for the hard scattering at leading order (LO) in QCD, and
  {\sc Pythia8}~\cite{Sjostrand:2006za,Sjostrand:2007gs} for QCD showering, fragmentation and hadronization. 
The {\sc MSTW2008lo68cl}~\cite{Martin:2009iq} parton distribution function (PDF) set was used. The key detector effects are parameterized based on the performance achieved by the LHC experiments; 
 the identification efficiency of photons, $b$-jets and $\tau$ leptons, and 
the energy resolution of photons and $b$-jets. 
In the future, these performance characteristics can serve as benchmarks for the design and simulation of future collider detectors. 

The Monte Carlo samples used in this study are stored in the {\sc ProMC} file format~\cite{Chekanov:2013mma,Chekanov:2013hfa} which is analyzable {\em via} {\sc root}~\cite{Brun199781}. 
 The anti-$k_T$ algorithm~\cite{Cacciari:2008gp} is used to reconstruct jets with the {\sc FastJet} package~\cite{Cacciari:2011ma} and  a distance parameter of 0.4.  
Stable particles (mean lifetimes greater than $3\cdot 10^{-11}$ seconds) are selected for jet clustering, and   neutrinos are excluded.  

Optimization of signal and background separation is achieved by employing the boosted decision tree (BDT) algorithm from the toolkit 
for multivariate analysis class of {\sc root}~\cite{Brun199781} and our results are given in terms of a corresponding gaussian significance ($N_\sigma$) for rejecting the background-only 
 hypothesis.
 
 \subsection{The $b\bar{b}\gamma\gamma$ Final State}
 \label{bbyy}

\begin{figure*}[t!]
\centering
\mbox{
\subfigure{
\includegraphics[scale=.365]{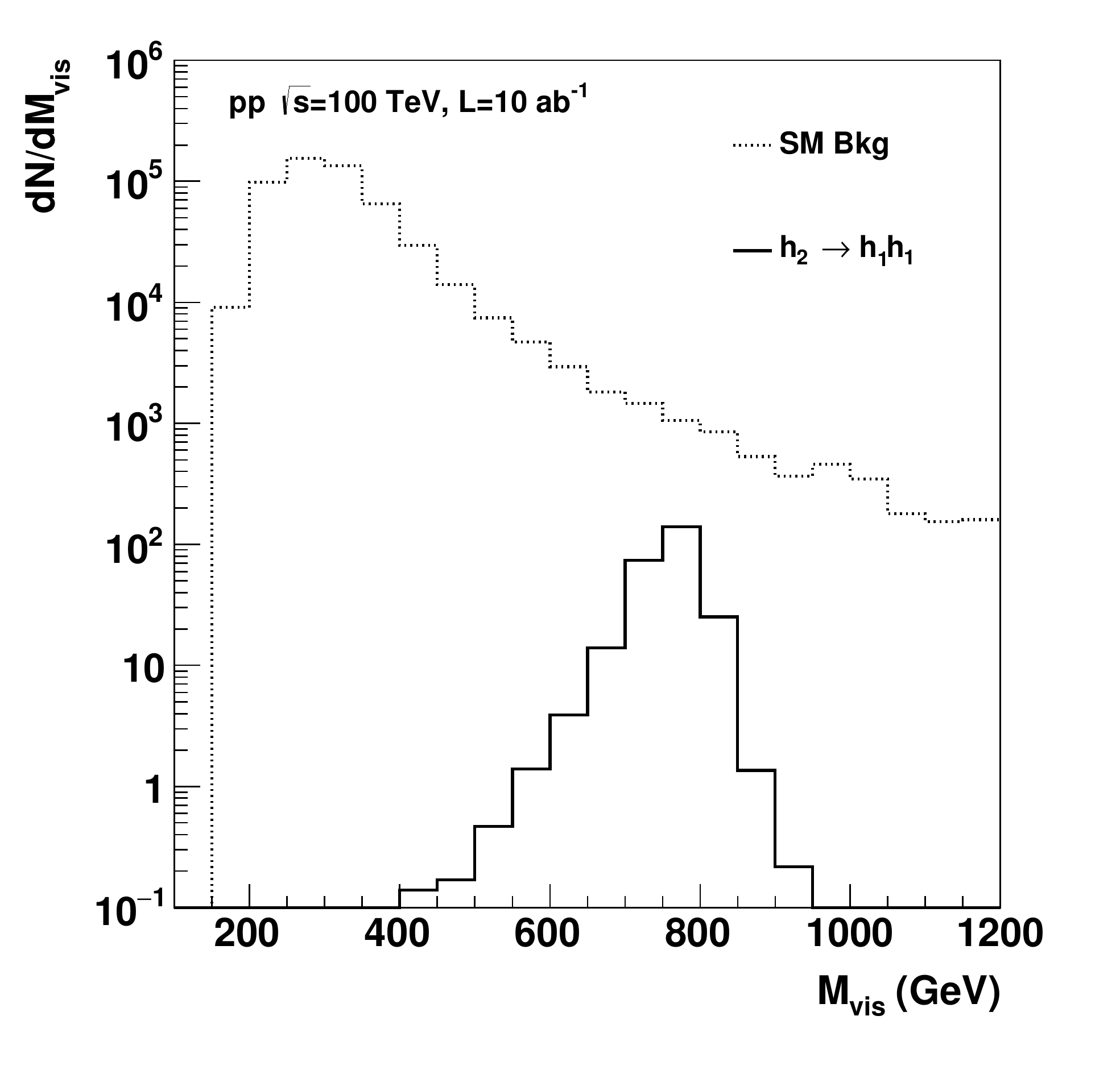}
}
\hspace{-.25in}\subfigure{
\includegraphics[scale=.365]{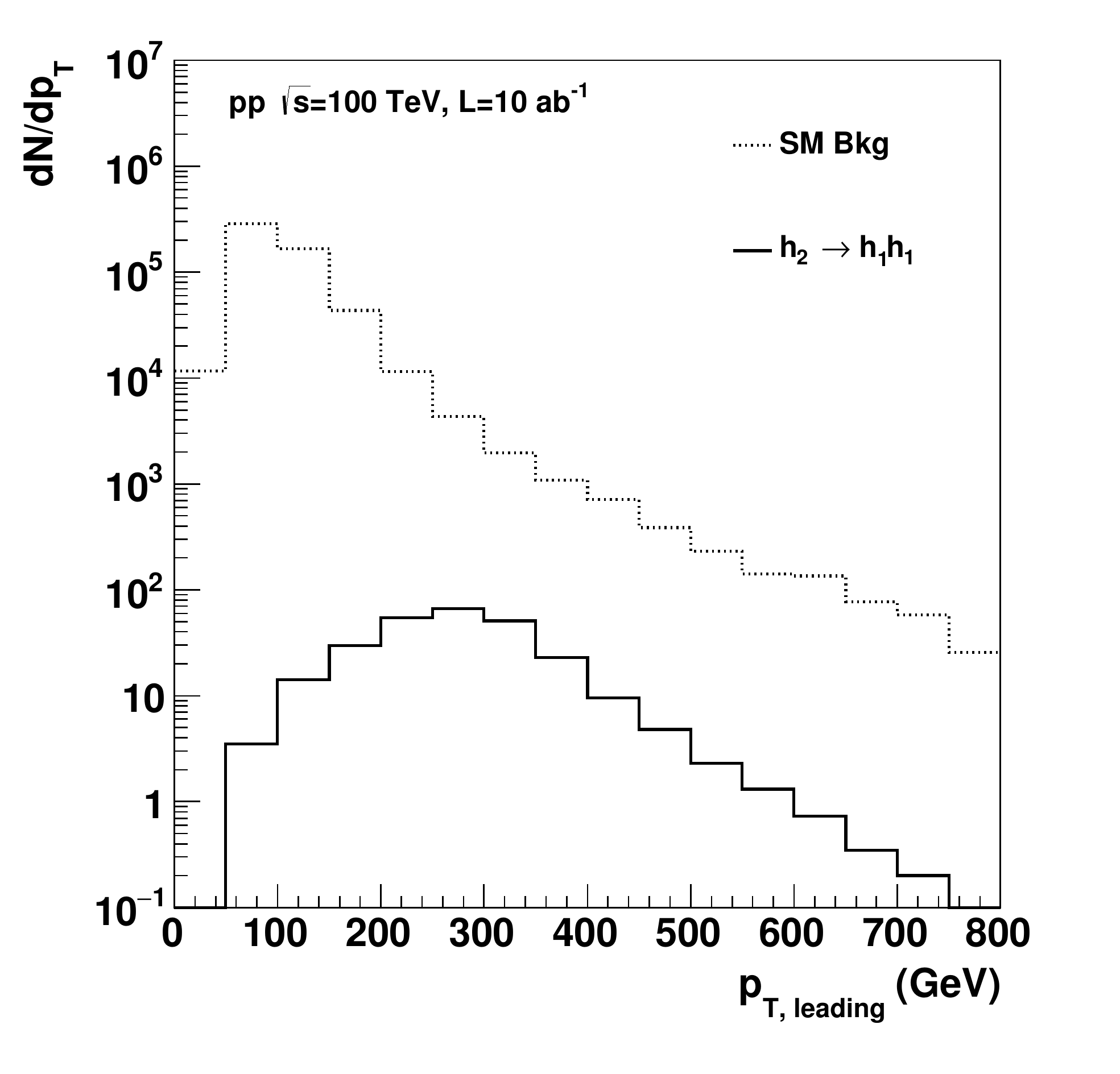}
}
}
\mbox{
\subfigure{
\includegraphics[scale=.365]{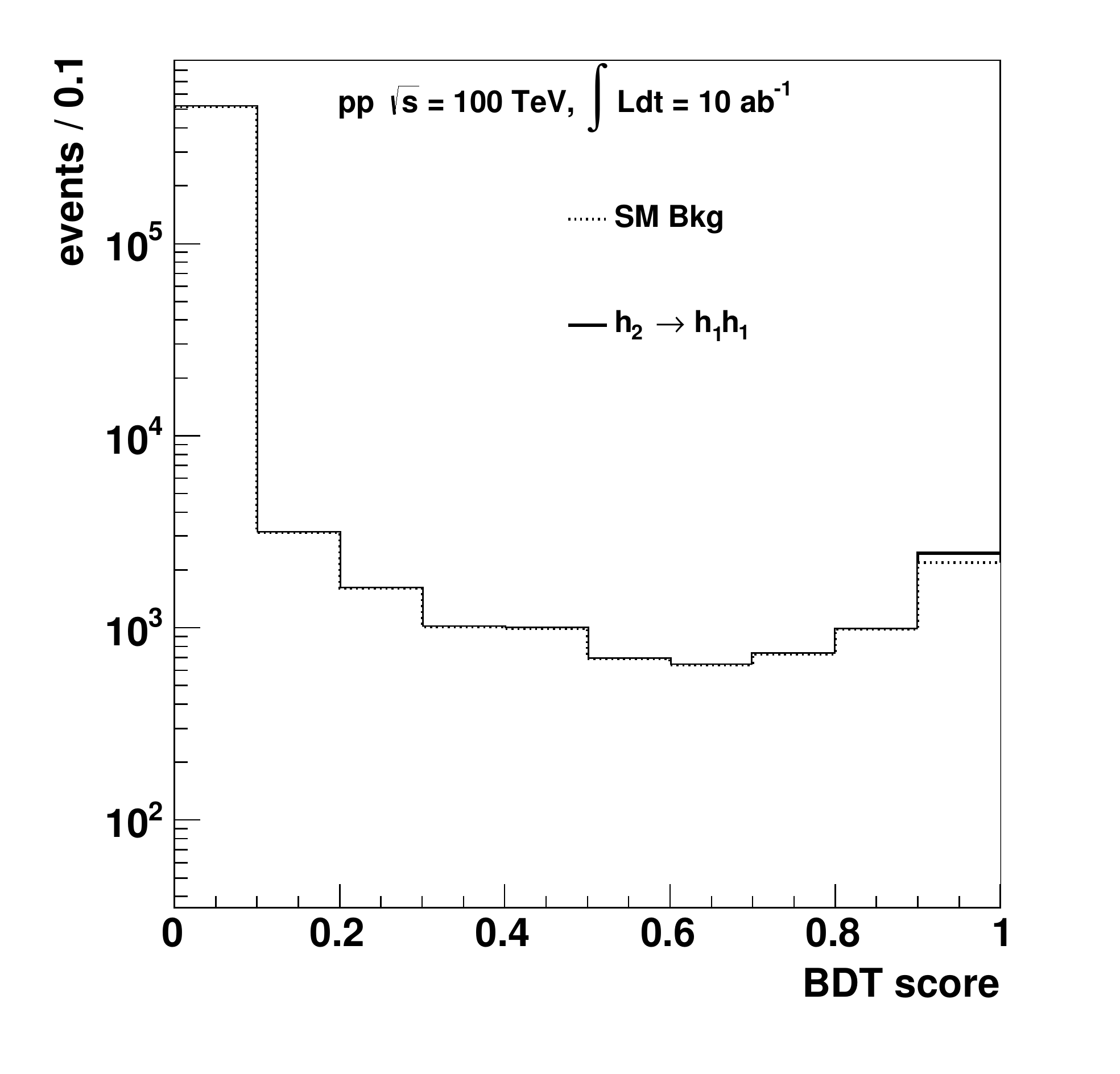}
}
\hspace{-.25in}\subfigure{
\includegraphics[scale=.365]{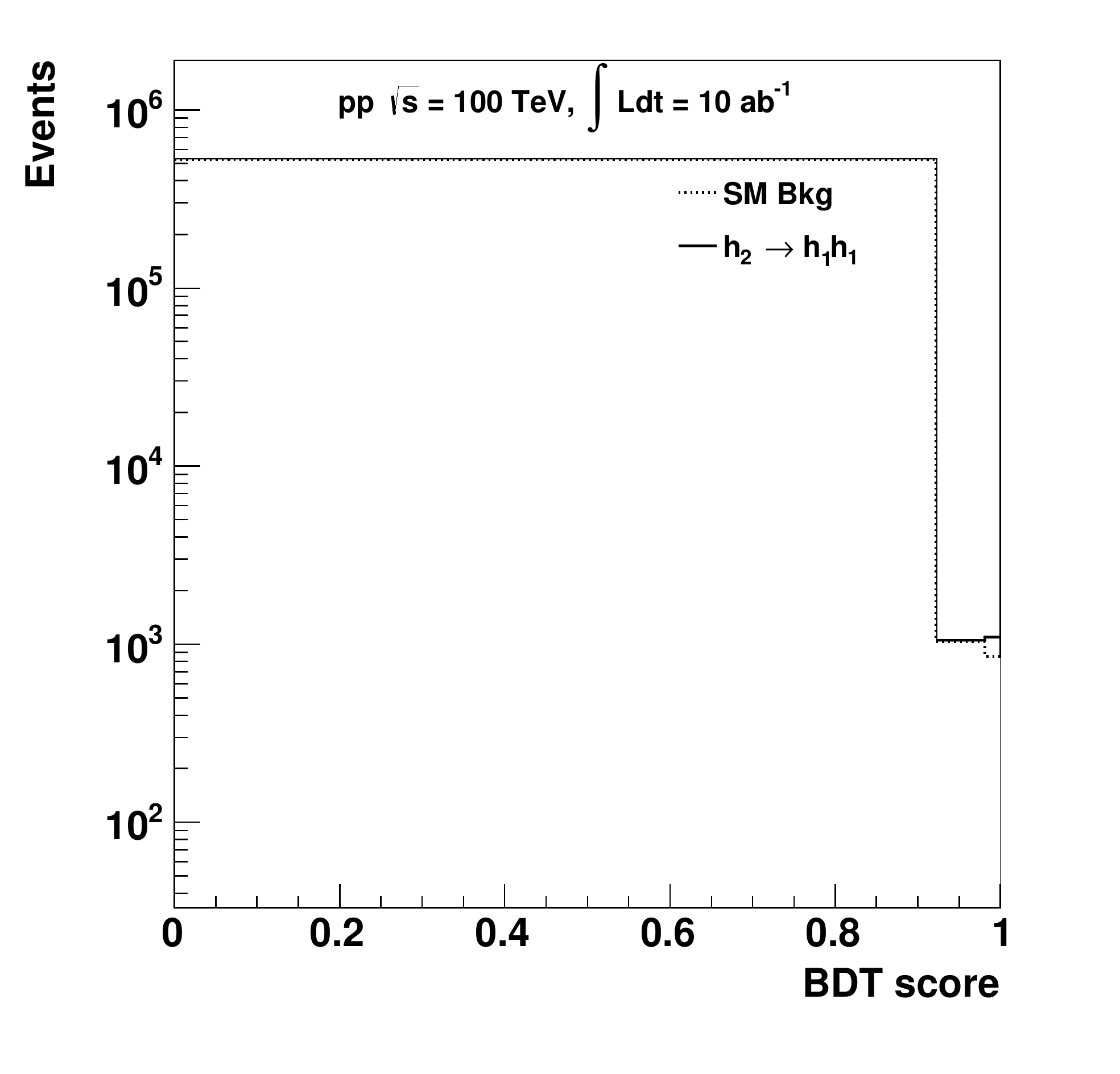}
}
}
\vspace{-5mm}
\caption{\small Signal and background distributions for the $b \bar{b} \gamma \gamma$ final state. The signal distributions correspond to  BM10$^\mathrm{max}$.
  The kinematic quantities shown are (top left) the invariant mass of the $b \bar{b} \gamma \gamma$
 system, and (top right) the $p_T$ of the leading particle from among the photons and the $b$-quarks. Also shown are the distributions of the BDT output with uniform binning (bottom left) and
 optimized binning (bottom right). }
\label{fig:bbaa_Dist}
\vspace{-2mm}
\end{figure*} 
 
 Here we present the analysis for the $b\bar{b}\gamma\gamma$ final state for the benchmarks from 
Tables~\ref{T1} and~\ref{T2}. 
 
For the backgrounds, all SM amplitudes contributing to the $\gamma \gamma b \bar{b}$  final state as well as the $\gamma \gamma t \bar{t}$
final state were included. This covers the following processes; $h (\to b \bar{b}) h (\to \gamma \gamma)$, 
 $b \bar{b} h (\to \gamma \gamma)$, $Z (\to b \bar{b}) h (\to \gamma \gamma)$, $t \bar{t} h (\to \gamma \gamma)$, and non-resonant
 $b \bar{b} \gamma \gamma $ and  $t \bar{t} \gamma \gamma $.  These processes have been  shown to be the dominant contributors of $b \bar{b} \gamma \gamma$ events 
 in the measurement of the trilinear Higgs self-coupling {\em via} Higgs pair 
 production~\cite{He:2015spf,Azatov:2015oxa}. It is also shown in~\cite{He:2015spf,Azatov:2015oxa} that backgrounds with jets misidentified as photons or $b$-jets contribute at most 25\% of the total background. Since the 
 mis-identification backgrounds  are non-resonant, we neglect these backgrounds with the understanding that they would degrade the sensitivity by $\cal{O}$(10)\%. Other neglected effects such 
 as higher-order QCD $k$-factors would compensate by enhancing the sensitivity by a similar amount. 
 
 The following cuts were applied at
 the generator-level: pseudorapidity $| \eta(\gamma) < 4| $, $| \eta(b) < 4| $, $p_T(\gamma) > 20$~GeV, $p_T(b) > 20$~GeV,  
$p_T^{\rm leading}(\gamma) > 40$~GeV, $p_T^{\rm leading}(b) > 40$~GeV, and the mass cuts $120 < m (\gamma \gamma) < 130$~GeV, $40 < m (b \bar{b}) < 200$~GeV.  
The visible cross sections with these requirements are presented in Tables~\ref{TbbaaMax} and~\ref{TbbaaMin} of Sec.~\ref{AppendixA}.
The efficiency for photon~\cite{Aad:2012tfa,Chatrchyan:2012ufa,1748-0221-10-08-P08010}
  and $b$-quark 
 identification~\cite{Aad:2015fna,Aad:2015ydr,1748-0221-8-04-P04013} was taken to be 75\% each~\cite{Aad:2008zzm,Chatrchyan:2008aa}. Photon energies were smeared by the electromagnetic 
calorimeter resolution 20\%/$\sqrt{E_T(\gamma)}$~$\oplus$~0.17\%~\cite{Aad:2012tfa,Aad:2014nim,Chatrchyan:2012ufa,1748-0221-10-08-P08010} and   
$b$-jet energies were smeared by the hadronic calorimeter resolution 100\%/$\sqrt{E_T(b)}$~\cite{Aad:2008zzm,Chatrchyan:2008aa,1748-0221-6-11-P11002,Aad:2012ag}. 

\begin{figure*}[t!]
\centering
\mbox{
\subfigure{
\includegraphics[scale=.65]{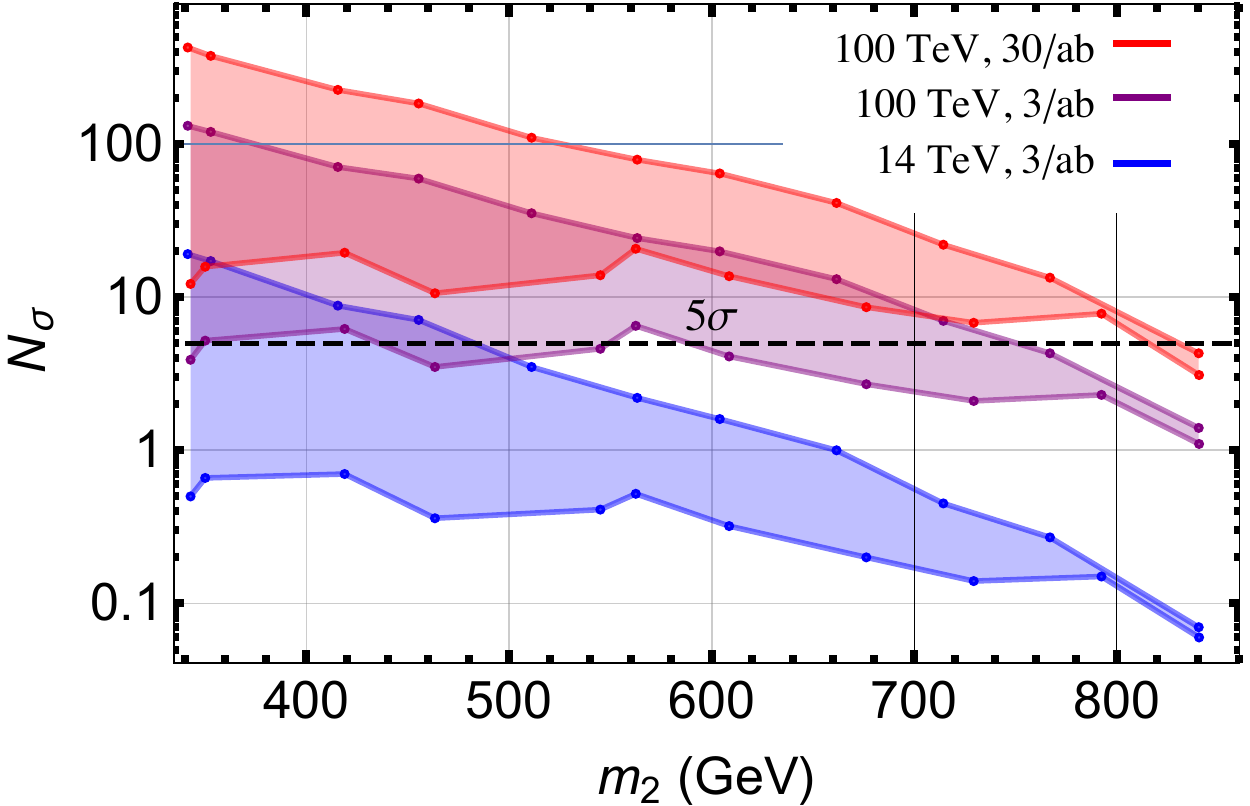}
}
\subfigure{
\includegraphics[scale=.65]{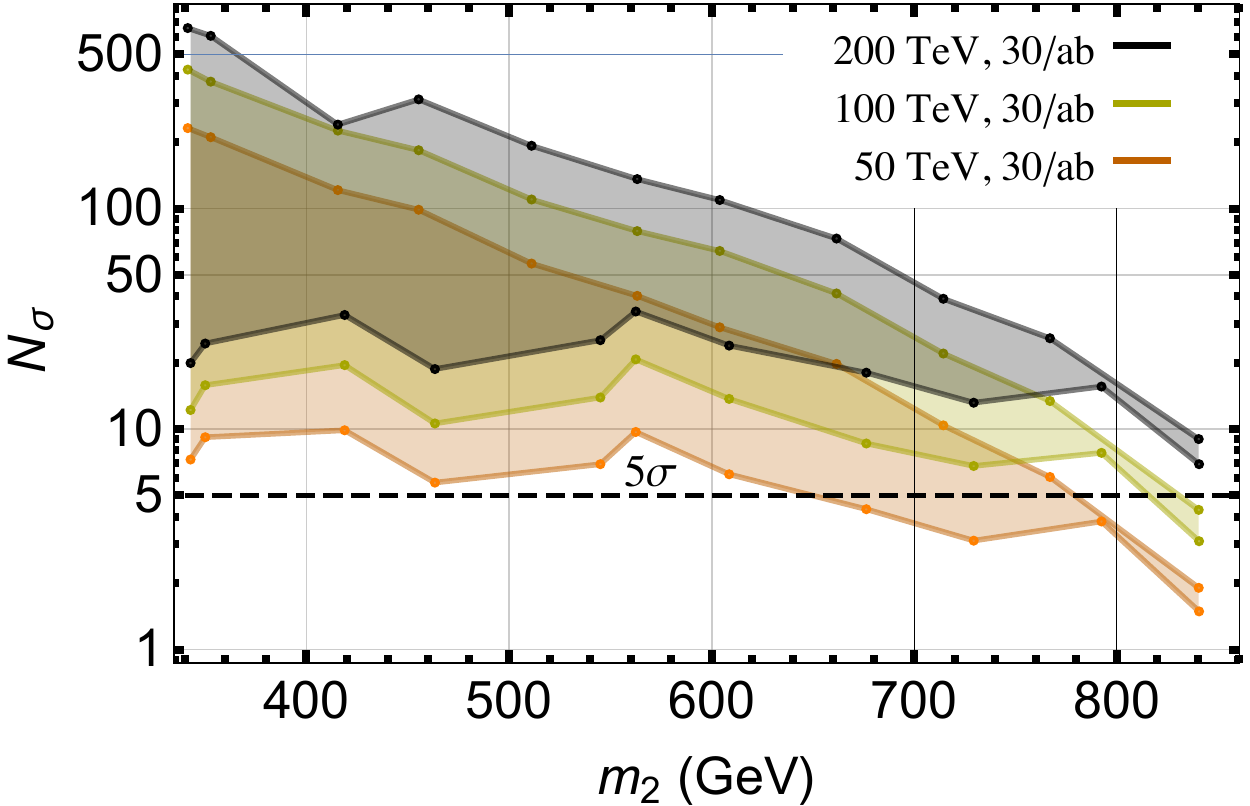}
}
}
\caption{\small The $N_\sigma$ gaussian significance for rejecting the background-only hypothesis, obtained using the $b\bar{b} \gamma \gamma$ final state, for each benchmark point. Different 
 collider scenarios of energy and integrated luminosity are compared. The vertical range corresponds to the maximum and minimum signal cross sections in the $h_2$ mass window.}
\label{BMplots}
\end{figure*}

A number of variables are computed whose distributions have different shapes for signal and background processes. The 
decay polar angle $\theta^*$ of the $h_2$ boson in its rest frame, with respect to the beam axis,  reflects its scalar nature {\em via} a uniform distribution in $\cos \theta^*$. 
The reconstructed invariant masses of the $h_1$ bosons and the $h_2$ boson show characteristic peaks compared to the smooth 
background distributions. The scalar sum $H_T = \Sigma |\vec{p_T}| $ of all visible objects   
excluding the $h_2$ decay products, is sensitive to additional jets in $t \bar{t}$ events.  
The magnitude of the vector sum $\met = | \Sigma \vec{p_T} |$ of all detected objects, which defines the missing transverse energy, can also be large in $t \bar{t}$ events.

 We also compare the distributions of the sphericity~\cite{Aad:2012np,Okorokov:2011mb} and the planarity~\cite{Aad:2012np,Okorokov:2011mb}
  of the event, as computed from the two photons and the two $b$-jets. In the rest frame of the reconstructed di-Higgs resonance, the sphericity tensor 
 is defined as $S^{\alpha \beta} = (\Sigma_i p_i^\alpha  p_i^\beta ) / (\Sigma_i |p_i|^2)$, where $\alpha, \beta$~ $\epsilon$~$ \{x,y,z\}$ and the sums run over the momentum 3-vectors of the 
 two photons and the two $b$-jets. The eigenvalues $\lambda_{1,2,3}$ of $S^{\alpha \beta}$ represent the fractional squared momenta along three orthogonal eigen-directions, 
  and satisfy the condition $\lambda_1 + \lambda_2 +\lambda_3 = 1$. Ranking the eigenvalues as $\lambda_1 \geq \lambda_2 \geq \lambda_3$, 
 the sphericity is defined as $\frac{3}{2} (\lambda_2 + \lambda_3)$ and the planarity is defined as $(\lambda_2 - \lambda_3)$. An event with large $\lambda_1$ has most of the momentum flowing 
 along the major axis of the ellipsoid and little momentum flowing in its two perpendicular directions, leading to a pencil-like event and a small value of sphericity. On the other hand, an event 
 with comparable momentum flow along at least two orthogonal directions leads to higher values of sphericity.  In the latter case, if $\lambda_3$ is small, the momentum flow is confined 
 to two orthogonal directions defining a plane, yielding a high value of planarity since there is little momentum flow out of the plane.   

The distributions of the reconstructed 4-body invariant mass of the $b \bar{b} \gamma \gamma$ system, M$_{\rm vis}$, 
 and the $p_T$ of the leading decay object (from amongst the two photons and the two  $b$-jets),  
 are shown in Fig.~\ref{fig:bbaa_Dist} for BM10$^\mathrm{max}$ after applying the following additional selection
requirements; $115 < m(\gamma \gamma) < 135$~GeV and $40 < m (b \bar{b}) < 200$~GeV. 
The other distributions mentioned above are shown in Fig.~\ref{fig:bbaa_Dist2} of Sec.~\ref{AppendixA}.
 
We combine the information in the following distributions: $p_T$ of the leading and sub-leading objects (photons and $b$-jets), 
the $\met$, the $H_T$, di-photon mass, $b \bar{b}$-mass,  the $\gamma \gamma b \bar{b}$ mass, the sphericity,
planarity and $\cos \theta^*$ using a BDT algorithm to separate the $h_2 \to h_1 h_1 \to \gamma \gamma b \bar{b}$ signal from the $\gamma \gamma b \bar{b}$ 
and $\gamma \gamma t \bar{t}$ backgrounds. The resulting distributions of the BDT score are shown in Fig.~\ref{fig:bbaa_Dist}. 
For optimal sensitivity, the distribution of the BDT score is binned such that each bin contributes the 
maximum poisson sensitivity, starting from the right edge of the histogram where the signal peaks. With this rebinned histogram (Fig.~\ref{fig:bbaa_Dist}), 
we quantify the discovery reach for the signal by computing the quantity $CL_b = P(Q<Q_{obs}|b)$, the probability for the test-statistic $Q$ to be smaller than the 
observed value given the background-only hypothesis. When $1 - CL_b < 2.8 \times 10^{-7}$
the background-only hypothesis is rejected at $5 \sigma$ significance. We convert this background fluctuation probability $1 - CL_b$ into the 
corresponding $N_\sigma$ gaussian significance and display them in Sec.~\ref{AppendixA} of the appendix (Tables~\ref{TbbaaMax} and~\ref{TbbaaMin}). 

Here we present these results in Fig.~\ref{BMplots}, where the shaded colored bands indicate the $N_\sigma$ ranges spanned by the BM$^\mathrm{max}$ and BM$^\mathrm{min}$ benchmark points. In  Fig.~\ref{BMplots} (left) we compare the reach of the HL-LHC with a $\sqrt{s}=100$ TeV $pp$ collider for two different values of integrated luminosity. Fig.~\ref{BMplots} (right) gives the comparison of three different prospective $pp$ collider energies for 30 ab$^{-1}$ of integrated luminosity. It is evident from Fig.~\ref{BMplots} (left) that for the $b{\bar b}\gamma\gamma$ channel, the HL-LHC could achieve discovery for a portion of the SFOEWPT-viable parameter space for  lower values of $m_2$, while a 100 TeV future collider with 30 ab$^{-1}$ could enable discovery over essentially all of the SFOEWPT-viable region. While the small region of phase space in the vicinity of the BM11 benchmark point ($m_2 > 820$~GeV) is below the $5 \sigma$ significance threshold with the $b \bar{b} \gamma \gamma$ channel alone,  the combination with another powerful channel using the $4 \tau$ final state renders this region also discoverable, as shown below.

 \subsection{The $4 \tau$ Final State}
 \label{4tau}
Here we present the analysis for the $4 \tau$ final state for the benchmarks from Tables~\ref{T1} and~\ref{T2}.

 Our background estimates include the SM processes producing four prompt $\tau$-leptons as these are expected to be the 
 dominant sources of backgrounds. Misidentification backgrounds will be suppressed to a negligible level as long as the future collider detectors achieve a $\tau$-identification efficiency
 and QCD jet rejection rate that are at least as high as the LHC experiments. For example, the ATLAS experiment reports 60\% efficiency for the identification of the hadronic decays of the 
 $\tau$-lepton,  with a QCD jet efficiency of 1-2\%~\cite{Aad:2014rga}. 
  Based on the construction of highly-granular electromagnetic calorimeters in the future, we assume an overall $\tau$-lepton identification efficiency of 75\%, inclusive of all decay channels.
 We emphasize this benchmark detector performance  for hadronic decays, which not only dominate the branching ratios but also provide the narrowest reconstructed Higgs and di-Higgs mass peaks due
 to the presence of fewer neutrinos in the final state.  
 
 It is shown in the $Z^\prime \to \tau \tau$  search~\cite{Aad:2015osa}, 
 where the transverse momenta ($p_T$) of the $\tau$-leptons are similar to our signal kinematics,
 that the dominant background in the double-hadronic mode arises from the $\gamma^*/Z \to \tau \tau$ Drell-Yan process.  The multijet and $W/Z$+jet backgrounds are a factor of 3-4 smaller. 
 The dominant background for our $HH \to 4 \tau$  search is the $ZZ \to 4 \tau$ process.  The diboson analysis for $ZV \to \ell \ell jj$~\cite{Aad:2014xka} 
 shows that  the $Z+$jets rate is about 20-50 times larger 
 than the $V V$ rate when the $Z \to \ell \ell$  and $V \to  jj$ masses are close to the $Z$ or $W$ boson masses. On the other hand, the hadronic $\tau$-lepton 
 selection suppresses QCD jets by a factor of 15 relative to prompt $\tau$-leptons. Thus, the requirement of two additional $\tau$'s will suppress the $Z+$jets background to a fraction of the $ZZ$ 
 background. Background from multi-jets, diboson+dijet and single-top+dijet processes will be suppressed even more strongly. Reference~\cite{Aad:2014xka} 
  also shows that the $t \bar{t}$ background is negligible
  when the bosons have high $p_T$. Thus, the inclusion of the SM $4 \tau$ background processes suffices for the estimation of the $h_1 h_1 \to 4 \tau$ resonance sensitivity. 

 \begin{figure*}[t!]
\centering
\mbox{
\subfigure{
\includegraphics[scale=.365]{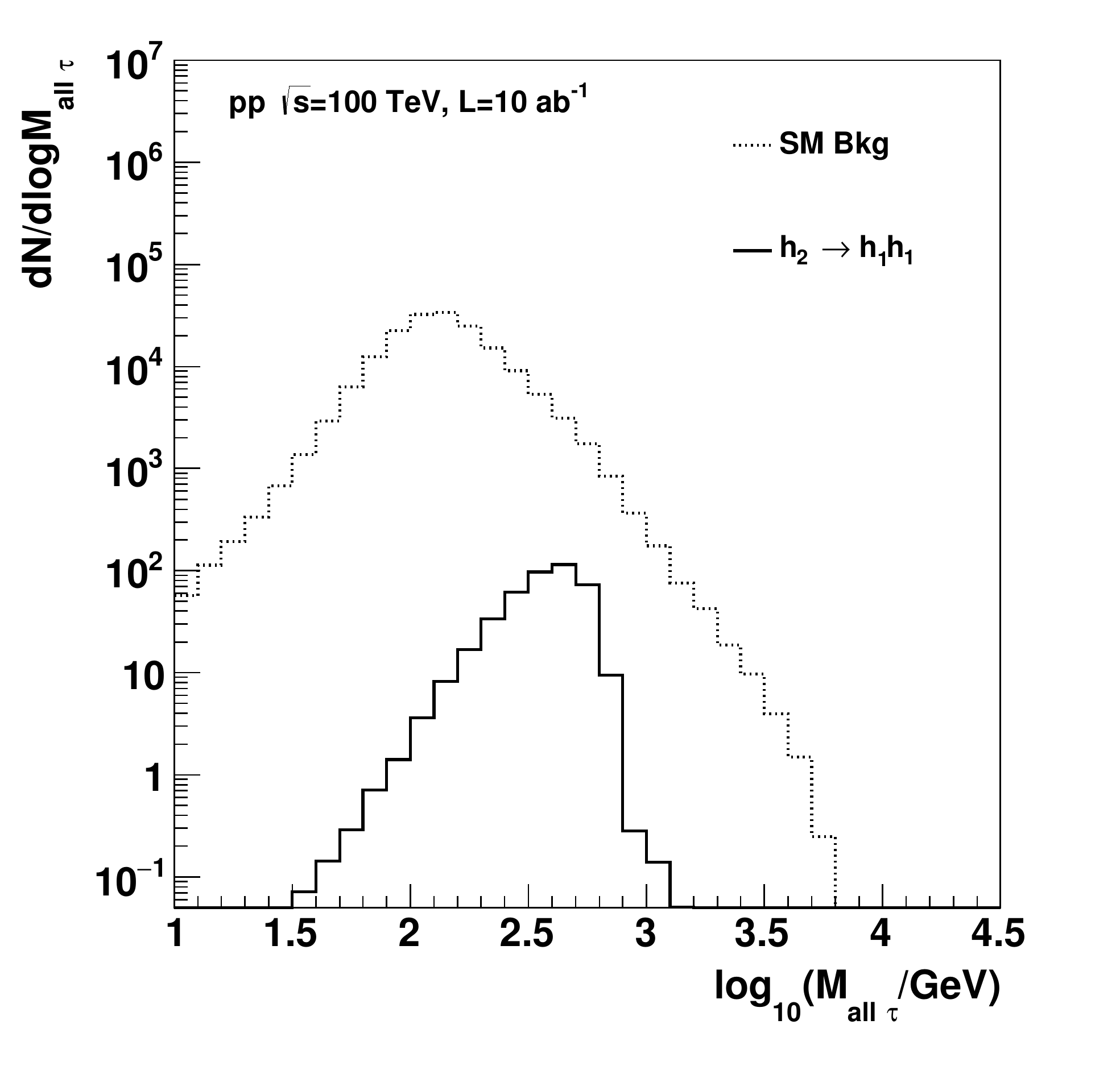}
}
\hspace{-.25in}\subfigure{
\includegraphics[scale=.365]{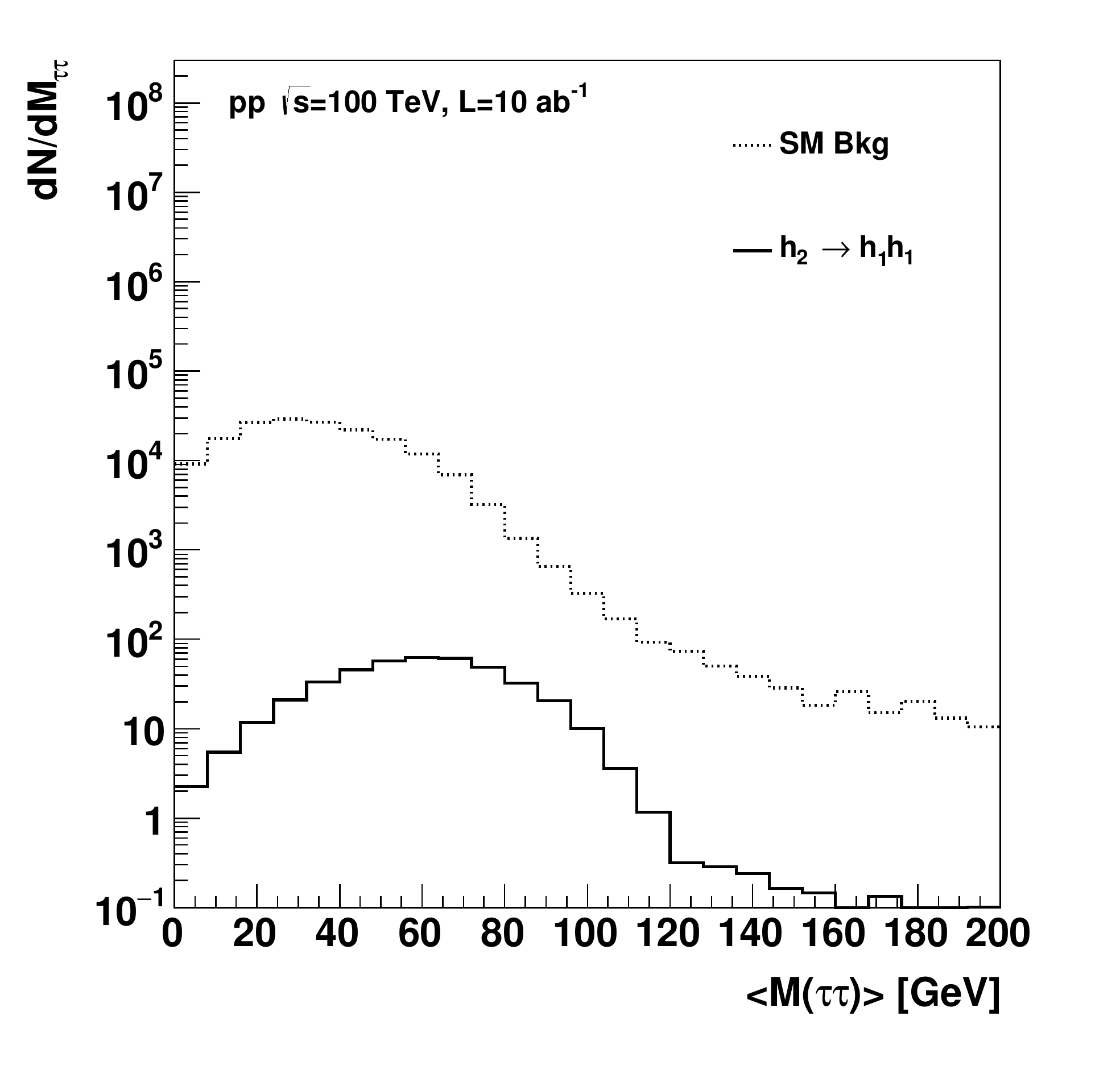}
}
}
\mbox{
\subfigure{
\includegraphics[scale=.365]{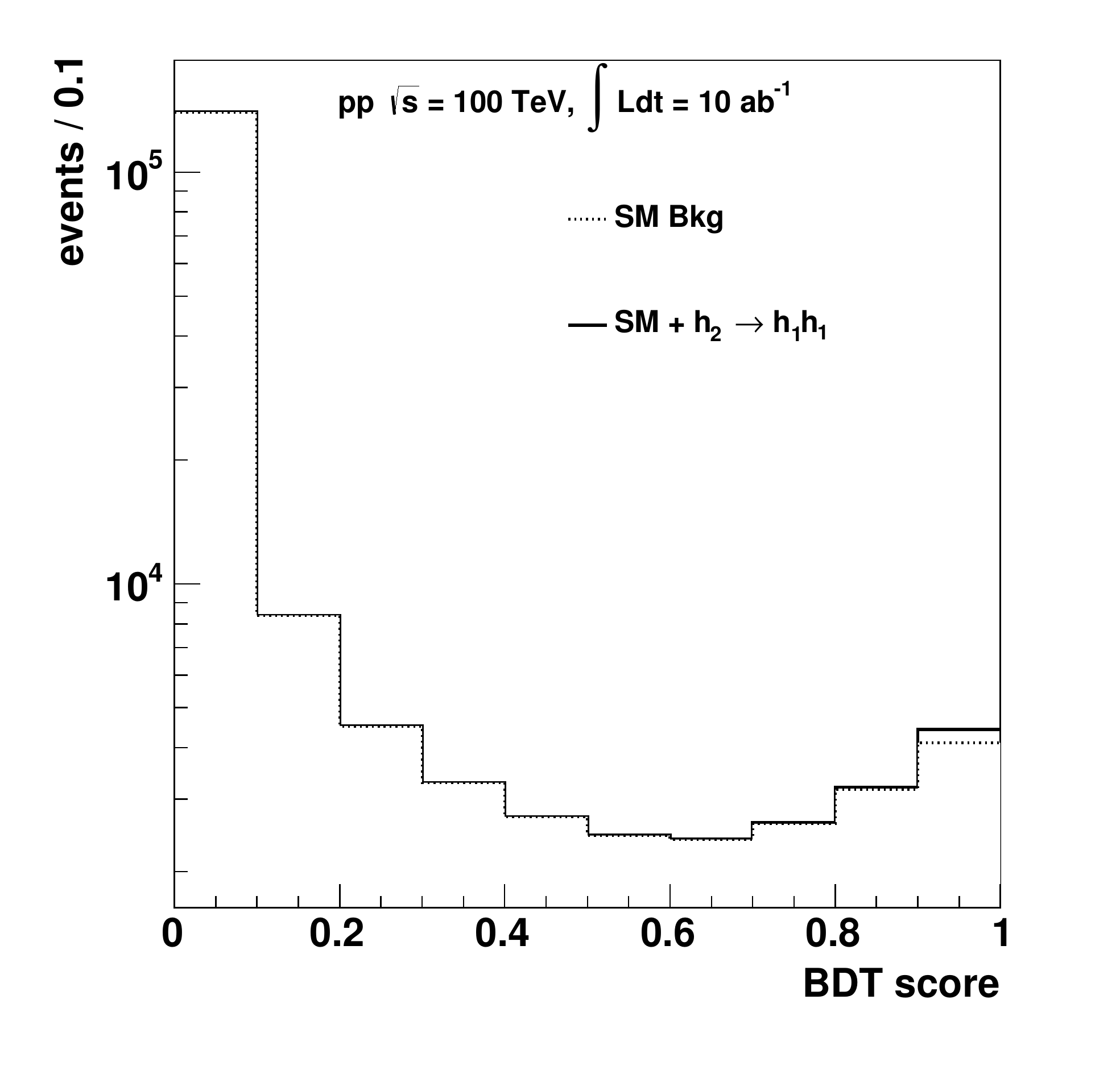}
}
\hspace{-.25in}\subfigure{
\includegraphics[scale=.365]{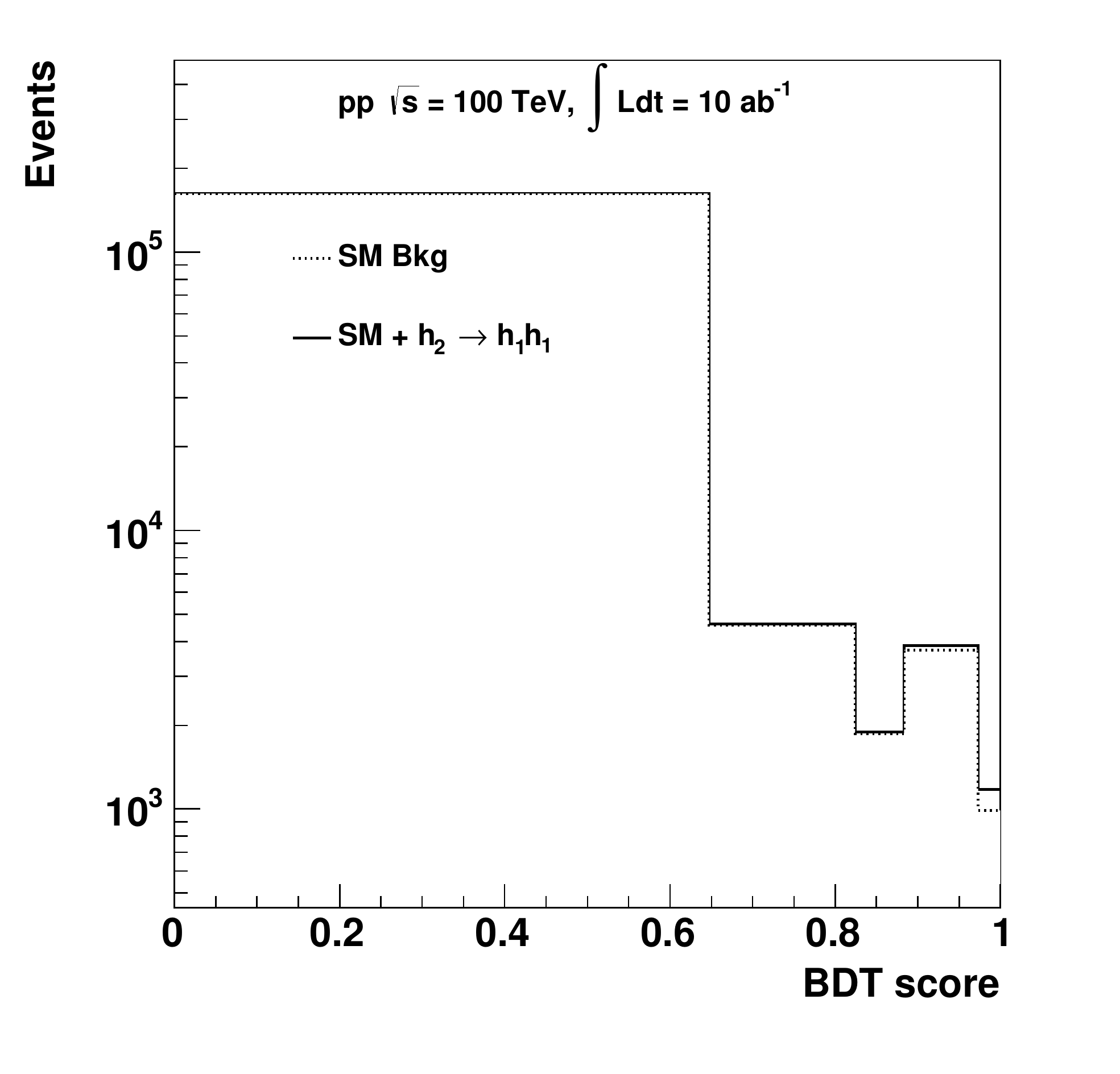}
}
}
\vspace{-6mm}
\caption{\small Signal and background distributions for the $4 \tau$ final state, where the signal corresponds to BM10$^\mathrm{max}$. 
 The kinematic quantities shown are (top left) the invariant mass of the $4 \tau$
 system, and (top right) the average  di-$\tau$ pair mass in the event. Also shown are the distributions of the BDT output with uniform binning (bottom left) and
 optimized binning (bottom right).}
\label{4tauDistros}
\vspace{-2mm}
\end{figure*}

\begin{figure*}[t!]
\centering
\hspace*{-.1in}
\mbox{
\subfigure{
\includegraphics[scale=.65]{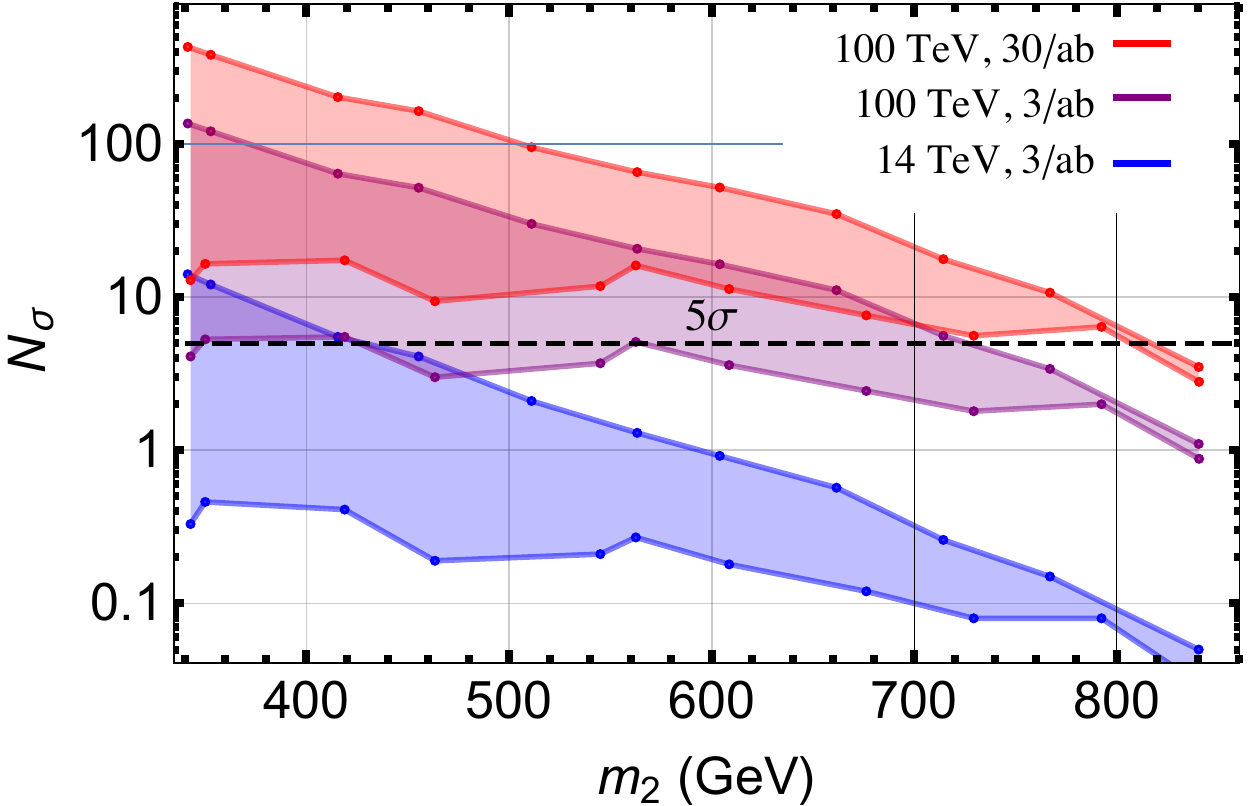}
}
\subfigure{
\includegraphics[scale=.65]{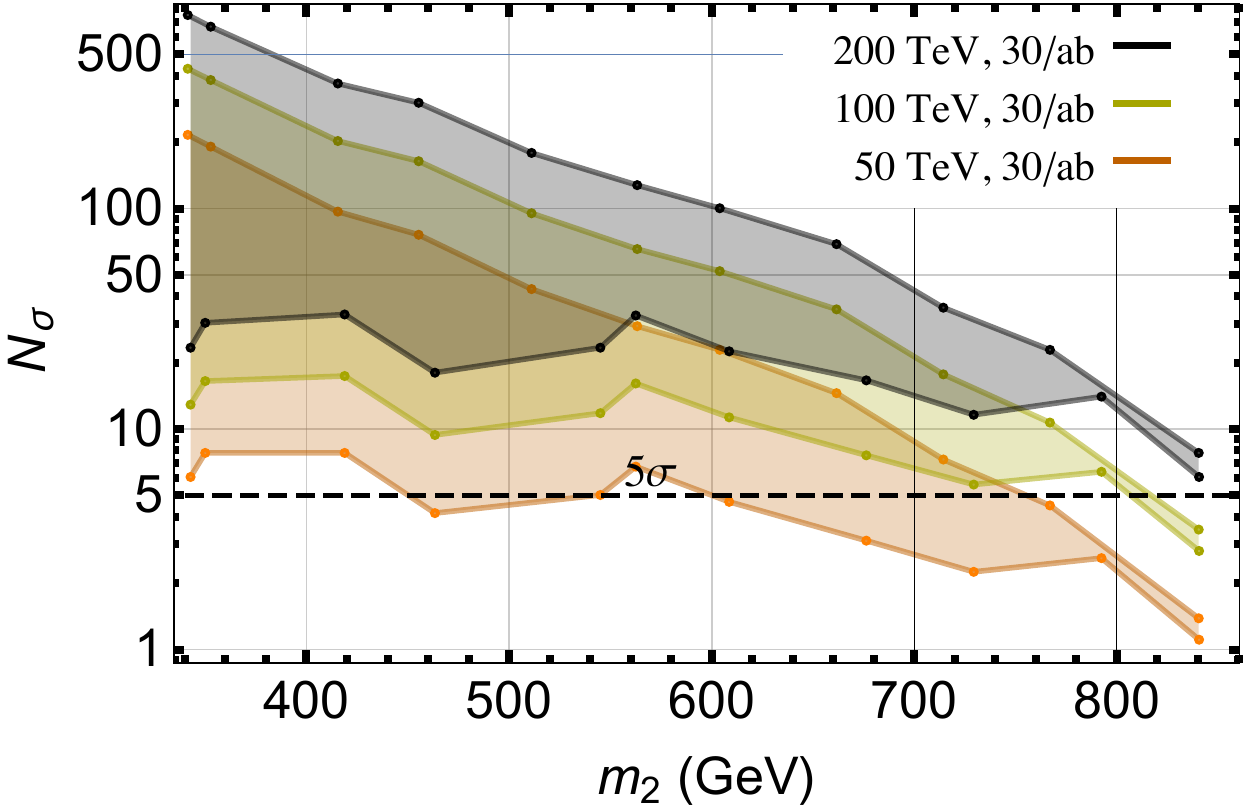}
}
}
\vspace{-2mm}
\caption{\small The $N_\sigma$ gaussian significance for rejecting the background-only hypothesis, obtained using the $4 \tau$ final state, for each benchmark point. Different 
 collider scenarios of energy and integrated luminosities are compared. The vertical range corresponds to the maximum and minimum signal cross sections in the $h_2$ mass window.}
\label{4tauBMplots}
\vspace{-3mm}
\end{figure*}

\begin{figure*}[t!]
\centering
\mbox{
\subfigure{
\includegraphics[scale=.65]{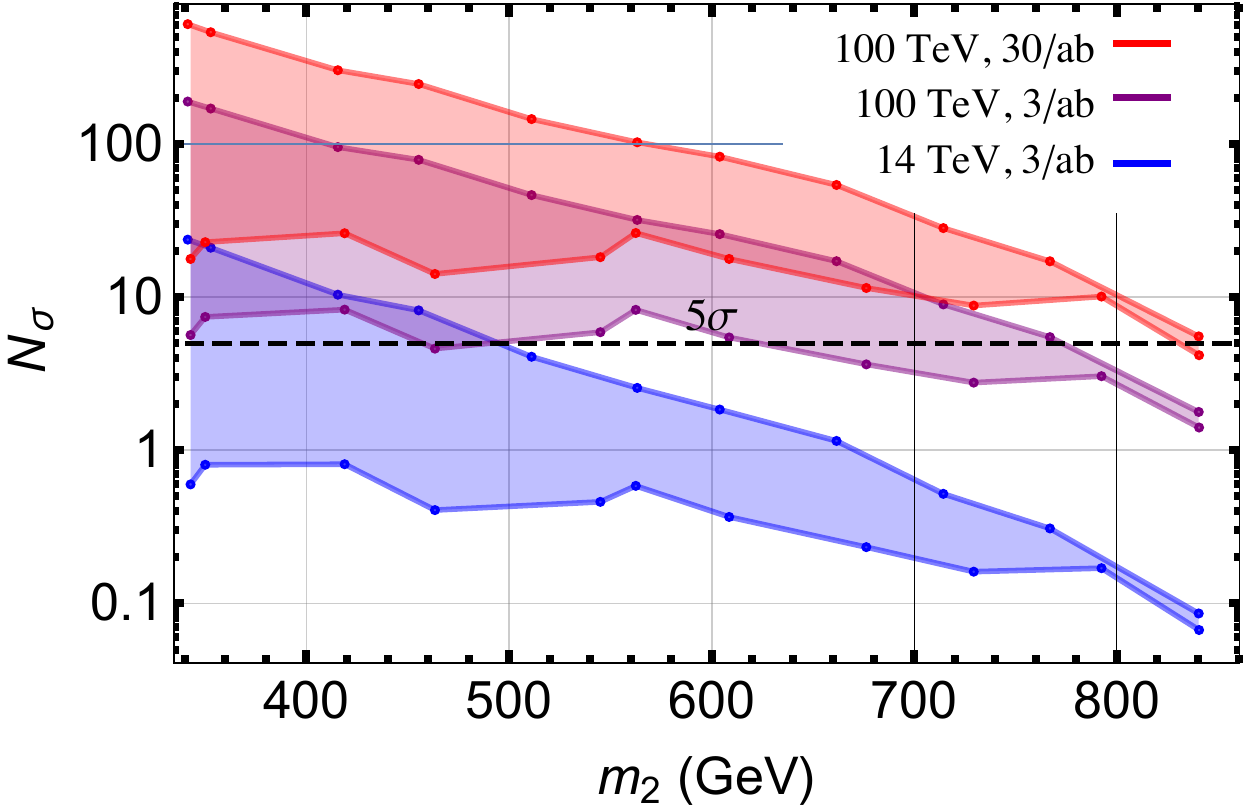}
}
\subfigure{
\includegraphics[scale=.65]{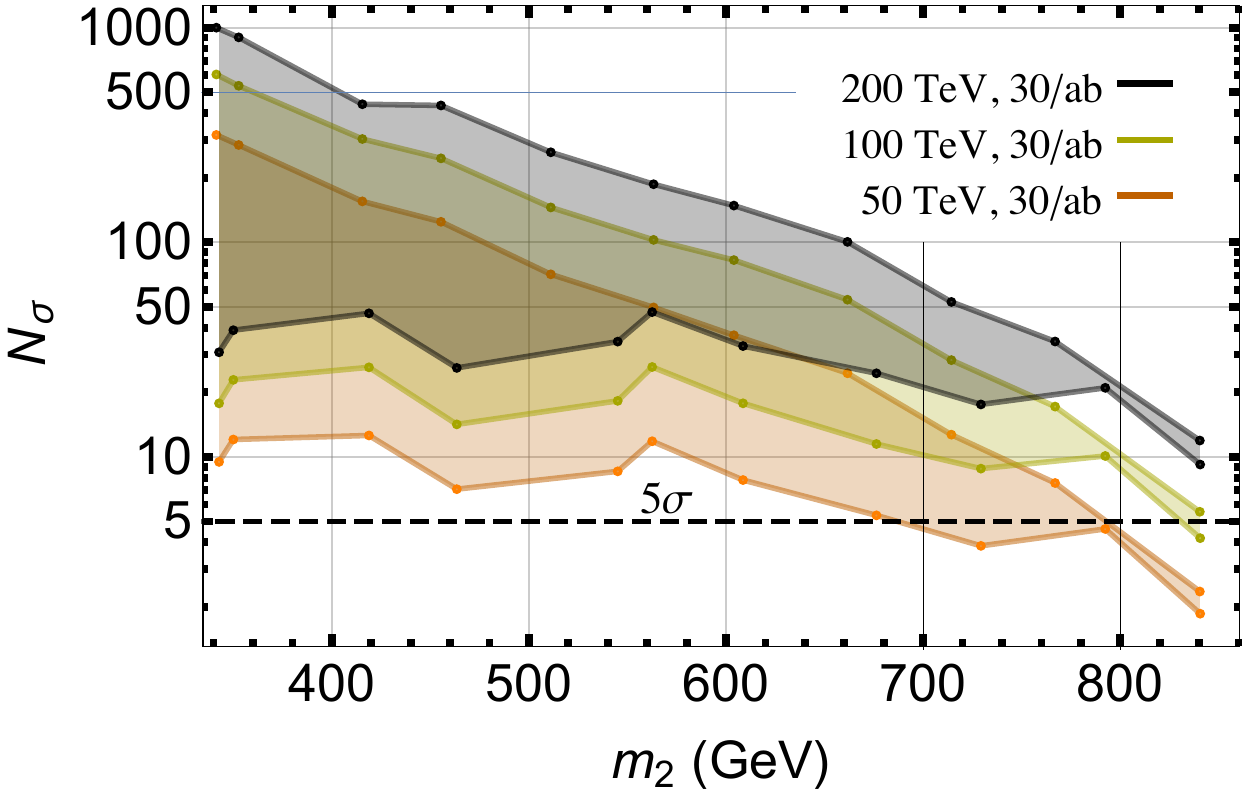}
}
}
\caption{\small The $N_\sigma$ gaussian significance for rejecting the background-only hypothesis, obtained using the combination of the $b \bar{b} \gamma \gamma$ and $4 \tau$ final states,
  for each benchmark point. Different 
 collider scenarios of energy and integrated luminosities are compared. The vertical range corresponds to the maximum and minimum signal cross sections in the $h_2$ mass window.}
\label{CombBMplots}
\vspace{4mm}
\end{figure*}

\begin{table*}[t!]
\caption{Combined results for the sensitivity $N_\sigma$ to $h_2 \to h_1 h_1$ production from the combination of $b \bar{b} \gamma \gamma$ and $4 \tau$ final states. The range ($N_\sigma^{\rm max}$ - $N_\sigma^{\rm min}$) indicates the variation in sensitivity that occurs when the signal cross section takes on its minimum and maximum allowed values within the range of parameter space that admits a SFOEWPT. }
 \vspace{5mm}
\begin{tabular}{|c|c|c|c|c|c|c|c|c|c|c|c|c|}
\hline
 & \multicolumn{2}{c}{14 TeV} & \multicolumn{2}{|c}{50 TeV} &  \multicolumn{6}{|c|}{100 TeV}  & \multicolumn{2}{|c|}{200 TeV}  \\
 \hline
 & \multicolumn{2}{c}{3 ab$^{-1}$} & \multicolumn{2}{|c}{30 ab$^{-1}$} & \multicolumn{2}{|c}{3 ab$^{-1}$} & \multicolumn{2}{|c}{10 ab$^{-1}$} & \multicolumn{2}{|c}{30 ab$^{-1}$} & \multicolumn{2}{|c|}{30 ab$^{-1}$} \\
\hline
 & $N_{\sigma}^{\rm min}$ & $N_{\sigma}^{\rm max}$ & $N_{\sigma}^{\rm min}$ & $N_{\sigma}^{\rm max}$ & $N_{\sigma}^{\rm min}$ & $N_{\sigma}^{\rm max}$ & $N_{\sigma}^{\rm min}$ & $N_{\sigma}^{\rm max}$ & $N_{\sigma}^{\rm min}$ & $N_{\sigma}^{\rm max}$ & $N_{\sigma}^{\rm min}$ & $N_{\sigma}^{\rm max}$ \\
\hline
B1  & 0.6  & 23.7 & 9.5 		& 316 		& 5.6 		& 189 		& 10.3 		& 347  			& 17.7 		& 606  			& 30.7 		& 1001 \\
\hline
B2 & 0.8 	 & 21.0  & 12.0 	& 284 		& 7.4 		& 170 		& 13.6 		& 313 			& 22.8 		& 537 			& 38.9 		& 902 \\
\hline
B3 & 0.81 & 10.4  & 12.6 	& 155 		& 8.3 		& 95.4 		& 15.2 		&  175 			& 26.1 		& 303 			& 46.6 		& 440 \\
\hline
B4 & 0.41 & 8.2 	  & 7.1 		& 124 		& 4.6 		& 78.8 		& 8.4 		& 143 			& 14.2 		& 246 			& 25.9 		& 434 \\
\hline
B5 & 0.46 & 4.1 	  & 8.5 		& 70.9 		& 5.9 		& 46.4 		& 10.9 		&  82.7 		& 18.2 		& 145 			& 34.4 		& 263 \\
\hline
B6 & 0.58 & 2.5 	  & 11.8 	& 49.7 		& 8.3 		& 31.9 		& 14.3 		& 58.1 			& 26.2 		& 103 			& 47.3 		& 186 \\
\hline
B7 & 0.36 & 1.8 	  & 7.8 		& 36.8 		& 5.4 		& 25.8 		& 10.1 		& 47.0 			& 17.7 		& 82.6 			& 32.8 		& 148 \\
\hline
B8 & 0.23 & 1.2 	  & 5.3 		& 24.5 		& 3.6 		& 17.2 		& 6.7 		& 30.4 			& 11.5 		& 54.0 			& 24.5 		& 100 \\
\hline
B9 & 0.16 & 0.52  & 3.8 	& 12.7 		& 2.7 		& 8.9 		& 5.0 		&  16.2 		& 8.8 		& 28.2 			& 17.5 		& 52.7 \\
\hline
B10 & 0.17 & 0.31& 4.6 	& 7.5 		& 3.0 		& 5.5 		&  5.8 		& 9.9 			& 10.1 		& 17.1 			& 21.0 		& 34.5 \\
\hline
B11 &0.07 & 0.08  & 1.8 	& 2.3 		& 1.4 		& 1.8 		& 2.4 		&  3.1 			& 4.2 		& 5.5 			& 9.2 		& 11.9 \\
\hline
\end{tabular}
\label{Tcombined}
\end{table*}

Signal and background processes are generated with requirements $p_T(\tau) > 20$~GeV, $p_T^{\rm leading}(\tau) > 40$~GeV and 
$| \eta (\tau) | < 4$. 
 The visible cross sections with these requirements are presented in Tables~\ref{T4tauMax} and~\ref{T4tauMin} of Sec.~\ref{AppendixB}.
Distributions of the signal (BM10$^\mathrm{max}$) 
 and background processes for the $4 \tau$ final state are shown in Fig.~\ref{4tauDistros}. Additional distributions are shown in Fig.~\ref{4tauDistrosAdditional}. 
As with the $\gamma \gamma b \bar{b}$ channel, we use a number of kinematic quantities computed with the $4 \tau$ final state as inputs to a BDT; the invariant mass of the four $\tau$-leptons, 
 the average di-$\tau$ mass (to distinguish between $Z \to \tau \tau$ and $h_1 \to \tau \tau$), $\cos \theta^*$, the sphericity and planarity of the event, $p_T$ of the leading and next-to-leading 
 $\tau$-leptons, the $\met$ and the $H_T$, to distinguish the $h_2 \to h_1 h_1 \to 4 \tau$ signal from the SM backgrounds. The resulting distribution of the BDT output is shown in
  Fig.~\ref{4tauDistros}. Again, we use the optimally-binned distribution of the BDT output to calculate the $N_\sigma$ gaussian significance of excluding the background-only hypothesis. We
  present the results from the $4 \tau$ channel in Tables~\ref{T4tauMax} and~\ref{T4tauMin} of the appendix in Sec~\ref{AppendixB} and, here, display these results in Fig.~\ref{4tauBMplots}. 

 \section{Combined Results }
 \label{Comb}

 The results presented in the previous section show that similar sensitivities to the $h_2 \to h_1 h_1$  process  are obtained from the $\gamma \gamma b \bar{b} $ and $4 \tau$ channels. 
 In the $\gamma \gamma b \bar{b} $ channel, the most discriminating variables are the di-photon and $b \bar{b}$ masses, the 4-body invariant mass, the event sphericity and planarity, and 
 the $p_T$ of the leading objects. The signal events have small sphericity and planarity compared to the backgrounds, due to the back-to-back decays of two $h_1$ bosons from a massive $h_2$ 
 boson. In the $4 \tau$ channel, the sphericity and the planarity values are also significantly smaller for signal than for backgrounds. The $4\tau$ mass and the average di-$\tau$ mass distributions 
 peak at higher mass for signal events, as do the $p_T$ for the leading $\tau$ leptons. The $H_T$ and missing $E_T$ variables also provide some discrimination as these variables have 
 higher values for signal events. 
  
 Here we present our final results, which compare the discovery potential for resonant di-Higgs pair production for various 
future collider scenarios in probing the xSM. The final results are obtained by combining the $N_\sigma$ sensitivities of the $\gamma \gamma b \bar{b} $ and $4 \tau$ channels. The 
 combination is performed by adding the respective $N_\sigma$ values in quadrature. The combined sensitivity is shown in Fig.~\ref{CombBMplots} and in Table~\ref{Tcombined}. 

 As mentioned earlier, the SFOEWPT-viable parameter space has a maximum $m_2 \sim 850$~GeV. 
 We find that with 30 ab$^{-1}$ of integrated luminosity, a 50 TeV $pp$ collider can achieve $5\sigma$ discovery of BM10 and lower $h_2$ masses, but falls short of discovering BM11. With the same
 integrated luminosity, a 100 TeV collider reaches the $5 \sigma$ threshold for BM11, and a 200 TeV collider achieves $10 \sigma$ sensitivity for the same. Thus, the higher 
 collider energies (or correspondingly
 higher integrated luminosities at lower energies) are needed to discover the $h_2 \to h_1 h_1$ process for $800 < m_2 < 850$~GeV, but the lower mass range can be discovered by lower energy colliders.
 
 We also note that a 100 TeV collider can discover up to BM7, and slices of the parameter space up to BM10, with 3 \abinv\ of integrated luminosity. Thus, increasing the integrated luminosity 
 to 30 \abinv\ enables the discovery in the $600 < m_2 < 850$~GeV mass range.  

\section{Conclusions}
\label{Conclusions}

Exploring the thermal history associated with EWSB is an important task for high energy physics. While EWSB in the SM with a 125 GeV Higgs boson is known to occur through a crossover transition, the addition of a single real gauge-singlet scalar to the scalar potential can significantly alter this picture. For a rather broad range of parameter choices in this simplest extension, EWSB may occur through a strong first order phase transition, thereby providing the out-of-equilibrium environment required for electroweak baryogenesis. In this context, it is interesting to determine the degree to which the LHC and prospective future high energy colliders might probe the SFOEWPT-viable parameter space of the \lq\lq xSM". 

In this study, we have attempted to address this question by considering parameter space regions that also allow for resonant di-Higgs production in $pp$ collisions. In doing so, we have identified a set of 22 SFOEWPT-viable benchmark parameter sets whose associated di-Higgs cross sections bracket the range of possible values in each of eleven 50 GeV-wide mass bins for the singlet-like scalar, $h_2$. Focusing on the HL-LHC and representative scenarios for higher-energy $pp$ colliders, we considered the corresponding reach of searches with the $b{\bar b}\gamma\gamma$ and $4\tau$ final states. We then asked: What would be the optimal center of mass energy and integrated luminosity for probing the SFOEWPT in the xSM? Our conclusions, to reprise the introductory discussion, are that:
\begin{itemize}
\item There exists interesting discovery potential for the HL-LHC for $m_2\lesssim 500$ GeV and exclusionary reach to somewhat higher masses
\item A 100 (200) TeV $pp$ collider with 30 ab$^{-1}$ of integrated luminosity could probe nearly all (all) of the SFOEWPT-viable parameter space.
\item A 50 TeV $pp$ collider with the same integrated luminosity would significantly extend the LHC reach, but would have a limited ability to probe the highest $m_2$ region.
\item Should future precision Higgs boson studies constrain the singlet-doublet mixing angle $| \theta | \lesssim 0.08$ (the currently projected limit from future 
 circular $e^+e^-$ colliders), there would still exist parameter choices for the xSM yielding a SFOEWPT. A future $pp$ collider as discussed here could discover the xSM even in this case.
 \item The gain in signal significance as a function of integrated luminosity $\mathcal{L}$ and collider energy can be summarized as follows. The signal significance increases with $\mathcal{L}$ as $\sqrt {\mathcal{L}}$ because the statistical fluctuations of both signal and background event yields are gaussian-distributed. The increase of collider energy from 50 TeV to 100 TeV increases the signal significance by a factor of 1.9 (2.3) at the low (high) mass benchmark points. For an increase of collider energy from 100 TeV to 200 TeV, the corresponding increase in signal significance is a factor of  1.7 (2.1).  Thus, a factor of four in integrated luminosity is roughly equivalent to a factor of two in collider energy, in terms of sensitivity to the $h_2 \to h_1 h_1$ process at a given $h_2$ mass.
\end{itemize}

It is important to take these conclusions somewhat impressionistically, as we have made a number of simplifying assumptions in order to paint the broad picture. 
\begin{itemize}
\item Theoretically, we have carried out a gauge-invariant analysis of the EWPT by working in the high-$T$ effective theory and omitting the $T=0$ Coleman-Weinberg contributions to the effective potential. Inclusion of the latter will, in general, yield additional parameter space regions consistent with a SFOEWPT. Moreover, the approximate criteria for baryon number preservation in Eq.~(\ref{eq:vctc}) is subject to additional theoretical uncertainties, some of which may be remedied with a future Monte Carlo study of the xSM phase transition dynamics.
\item Experimentally, we have considered a detector performance similar to the LHC detectors but extended up to $| \eta | < 4$, which is the goal for future collider detector design. We find that
 the $b{\bar b}\gamma\gamma$ and $4\tau$ final states have equal sensitivities for probing resonant di-Higgs production. The conclusions presented above are based on the combination of the 
 sensitivities from these channels, and emphasize the importance of achieving high acceptance and efficiency for photons, $b$-jets and hadronic $\tau$-leptons.  
\end{itemize}

With these considerations, we believe that our study provides a reasonable guide to what may be possible with a higher 
energy $pp$ collider and how it may compare with the HL-LHC. Moreover, for both the LHC and a future collider, the reach may be enhanced by considering other final states not studied here. For these reasons, the opportunities with a 100 TeV collider appear to be quite promising. Additional investigation of the energy frontier as a probe of the EWPT, thus, appears well worth the effort.

\begin{acknowledgments}
The work of AVK was supported by the Fermi National Accelerator Laboratory. 
Fermilab is operated by Fermi Research Alliance, LLC, under Contract No. DE-AC02-07CH11359 with the United States Department of Energy. 
The work of MJRM and PW was supported in part by U.S. Department of Energy contracts DE-SC0011095 and by 
the National Science Foundation under Grant No. NSF PHY11-25915. JMN is supported by the People Programme (Marie Curie Actions) of 
the European Union Seventh Framework Programme (FP7/2007-2013) under REA grant agreement PIEF-GA-2013-625809.
\end{acknowledgments}

\newpage

\section{Appendix A: $b\bar{b}\gamma\gamma$ Analysis}
\label{AppendixA}

In this section, we display the remaining kinematic distributions used in our BDT analysis which were not included in the main text of section~\ref{bbyy}. We also include here tables of cross sections and $N_\sigma$ results from our BDT analysis for the sets of benchmark points yielding the maximum and minimum signal cross sections.

\begin{widetext}
\onecolumngrid

\begin{figure*}[h!]
\centering
\mbox{
\subfigure{
\includegraphics[scale=.23]{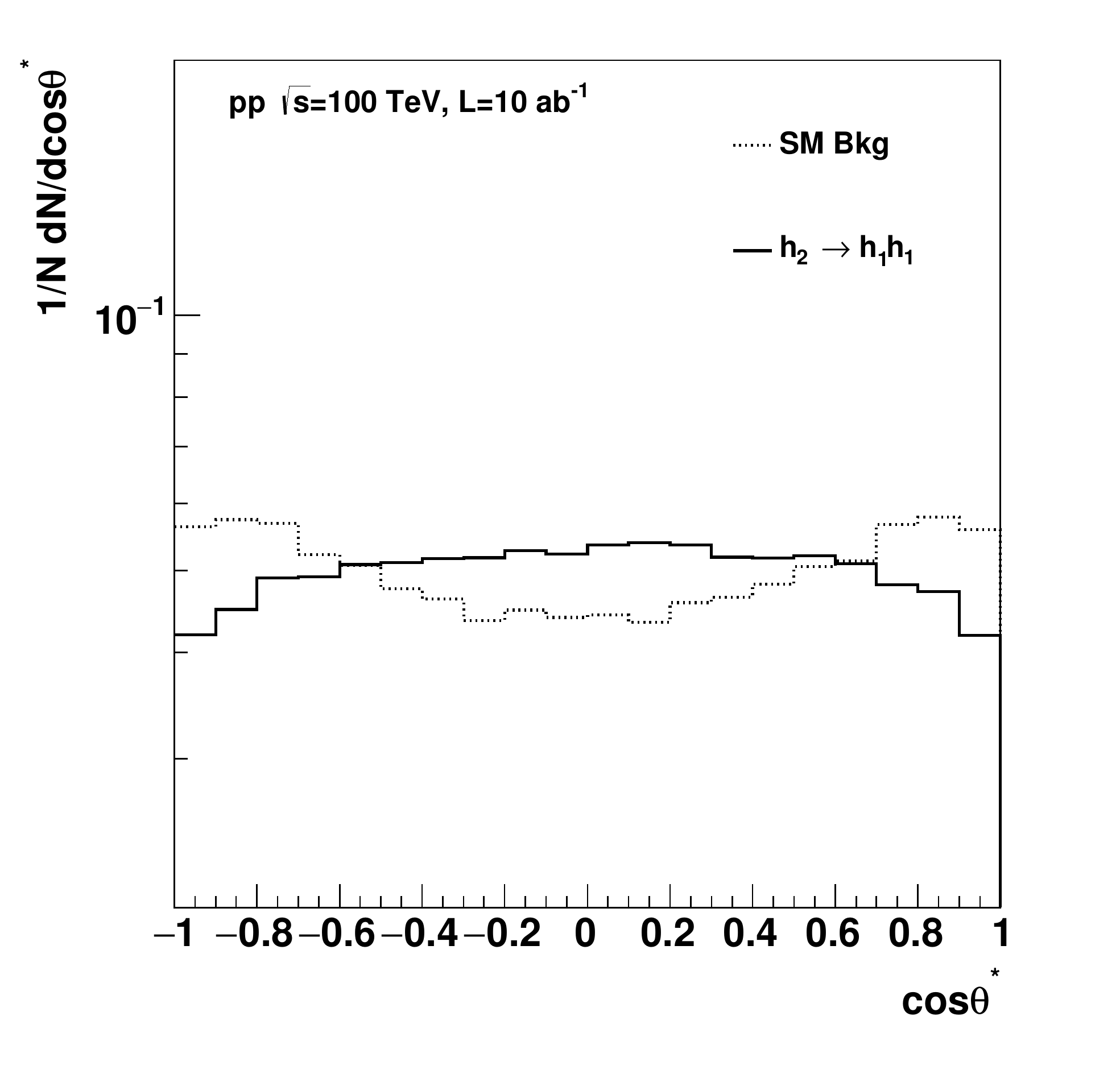}
}
\hspace{-.25in}\subfigure{
\includegraphics[scale=.23]{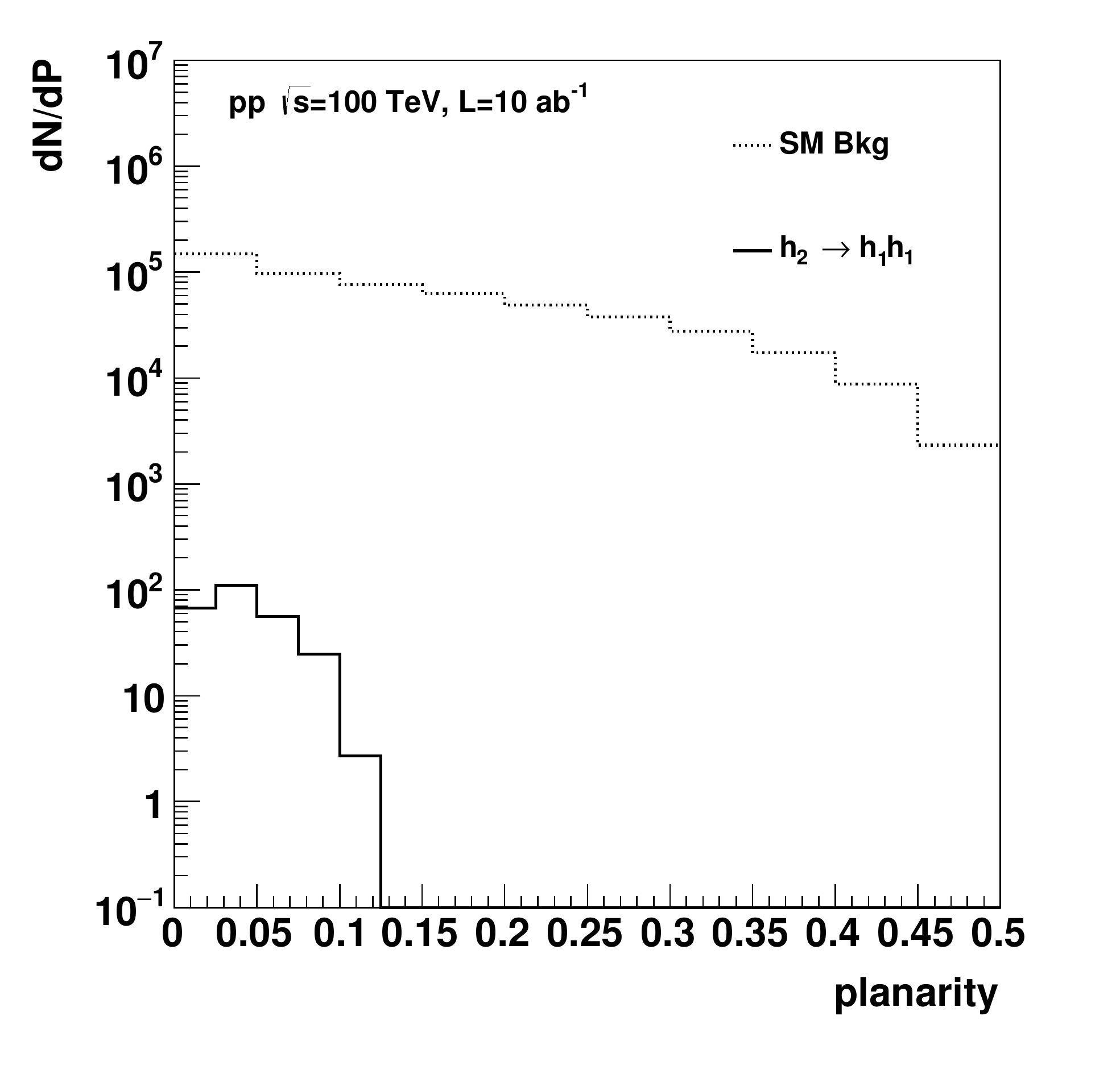}
}
\hspace{-.25in}\subfigure{
\includegraphics[scale=.23]{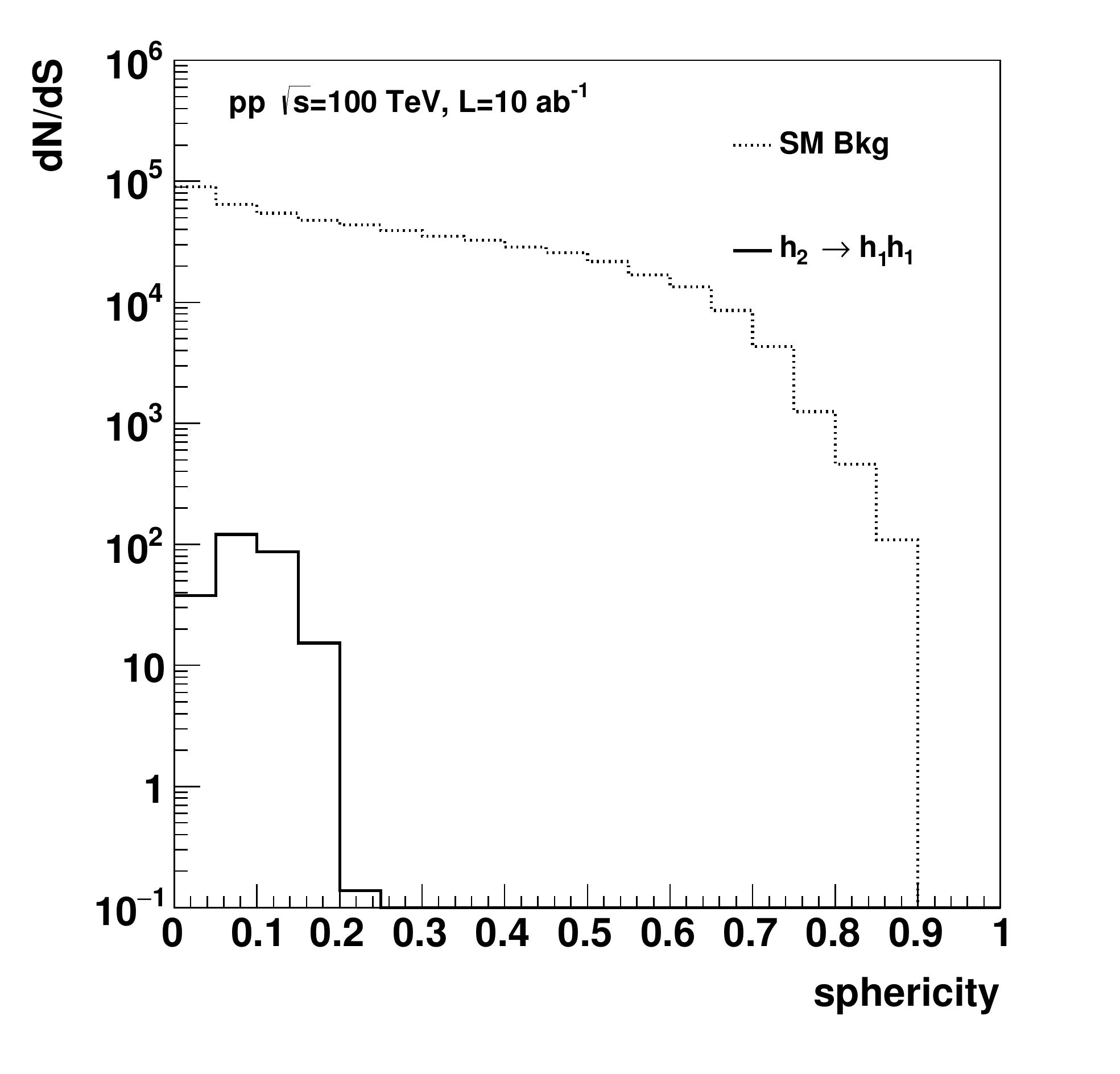}
}
\hspace{-.25in}\subfigure{
\includegraphics[scale=.23]{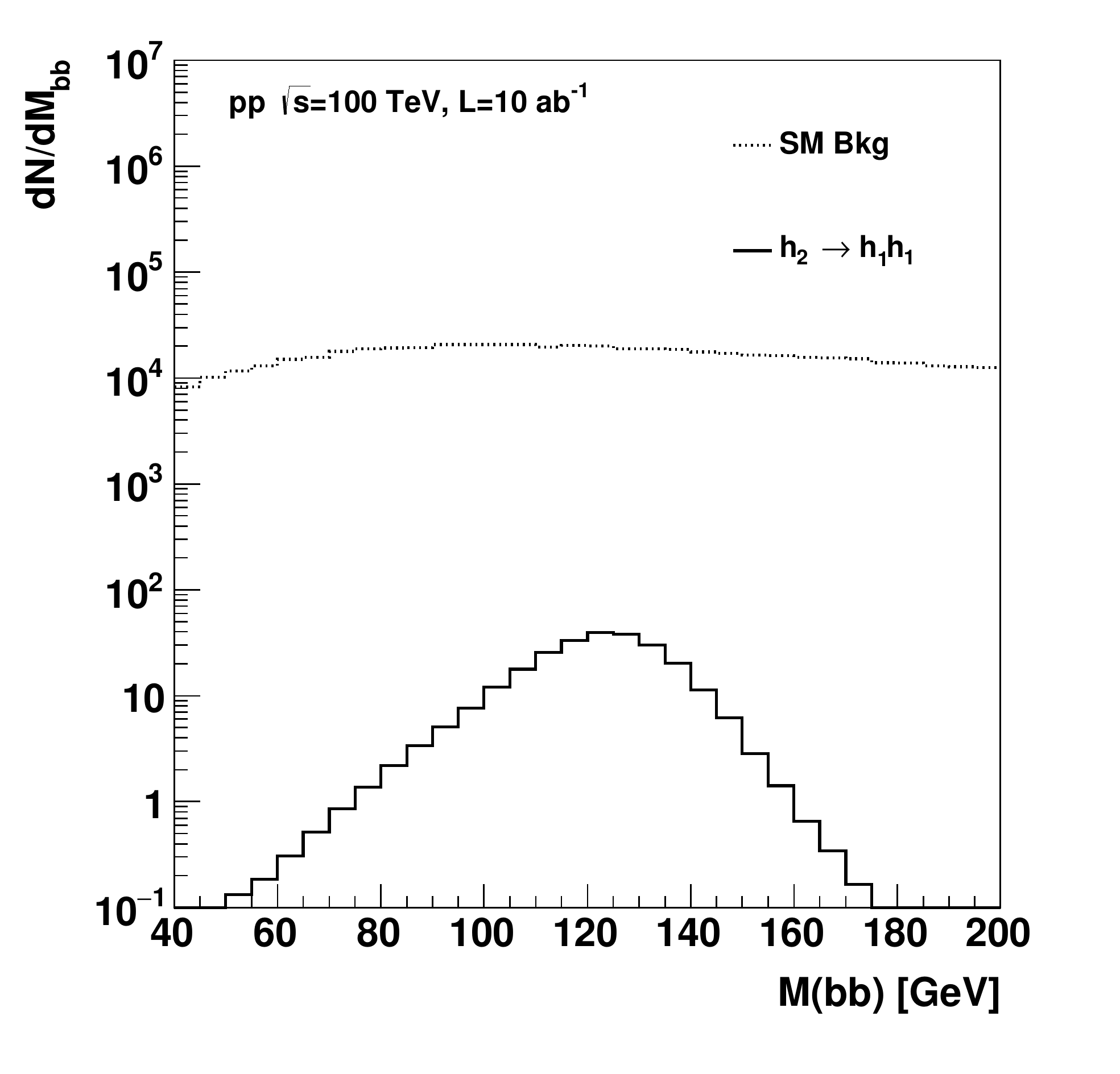}
}
}
\mbox{
\subfigure{
\includegraphics[scale=.23]{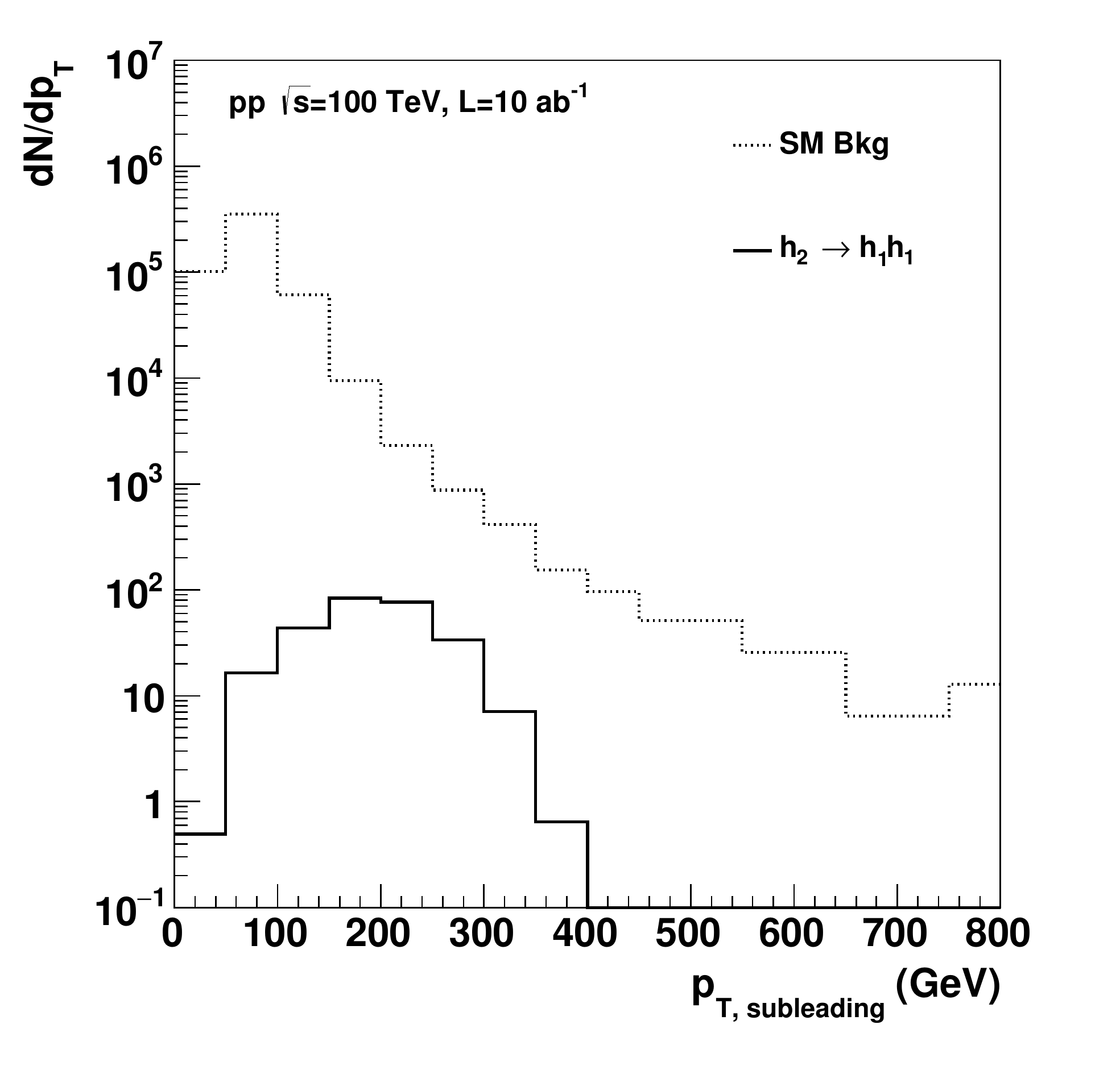}
}
\hspace{-.25in}\subfigure{
\includegraphics[scale=.23]{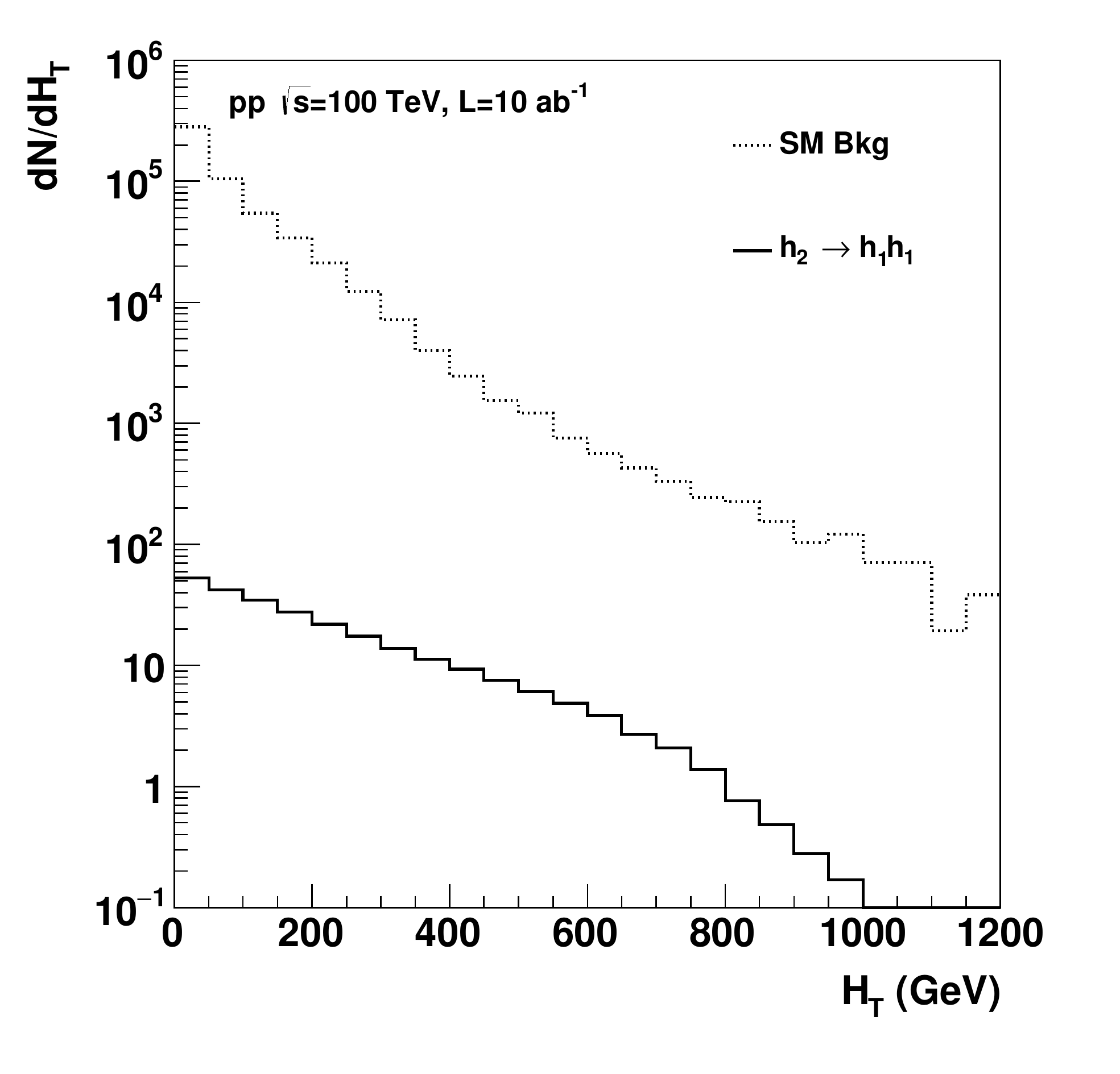}
}
\hspace{-.25in}\subfigure{
\includegraphics[scale=.23]{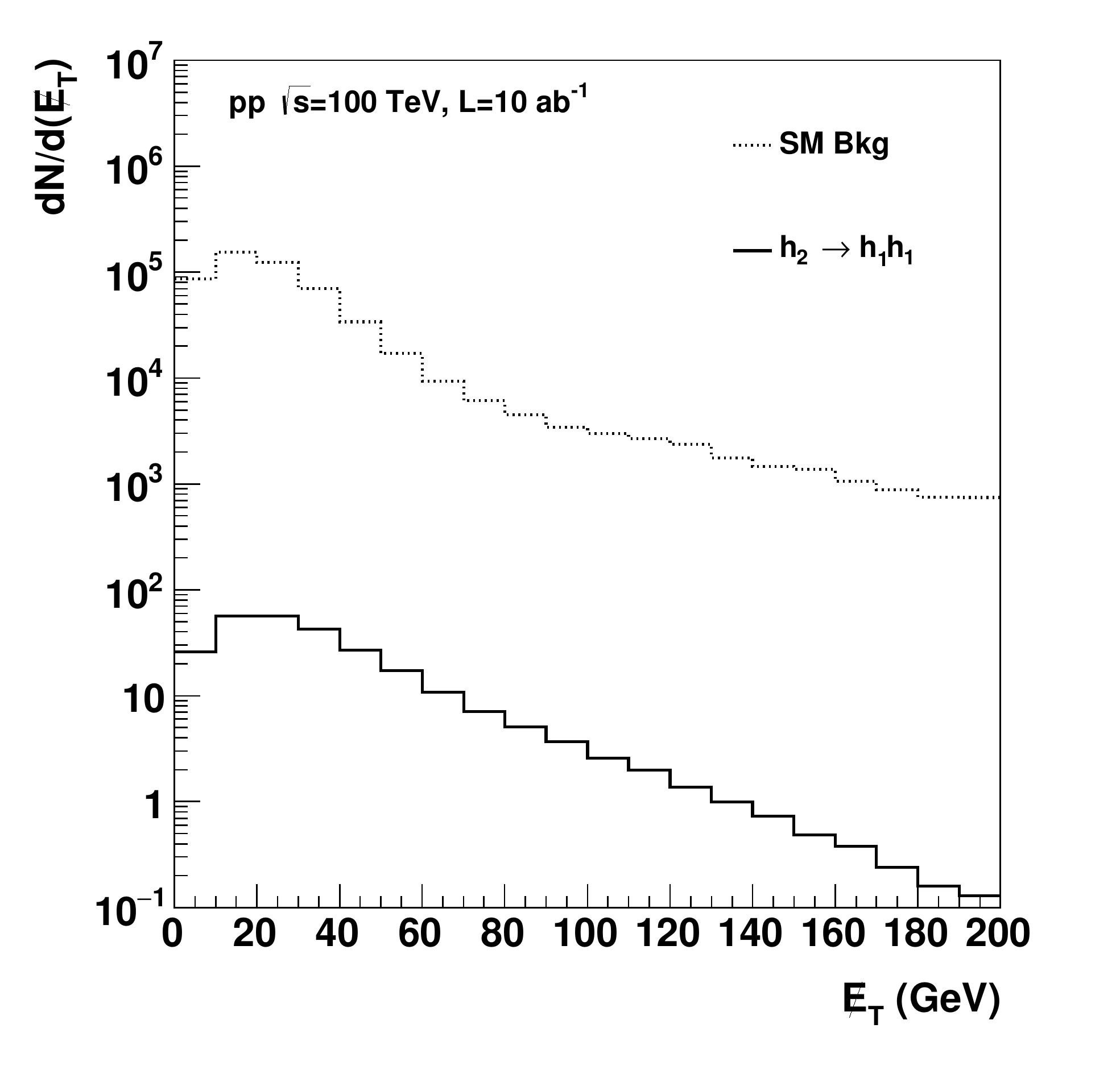}
}
\hspace{-.25in}\subfigure{
\includegraphics[scale=.23]{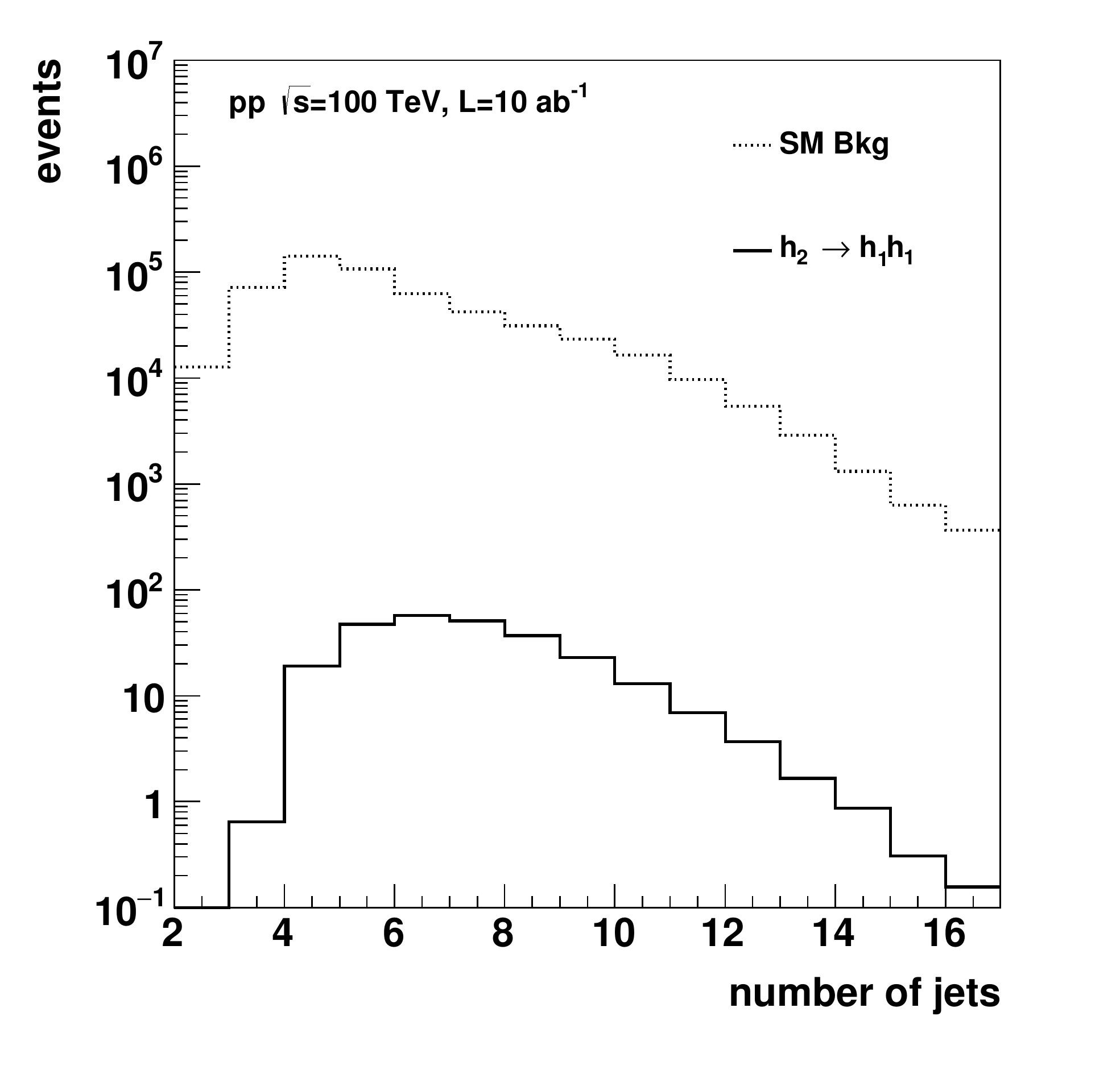}
}
}
\caption{Additional kinematics distributions for $b \bar{b} \gamma \gamma$ final state, used as inputs to the BDT. The signal  distributions correspond to BM10$^\mathrm{max}$. }
\label{fig:bbaa_Dist2}
\end{figure*}

\begin{table*}[h!]
\caption{Cross sections and BDT analysis results for $b \bar{b} \gamma \gamma$ final state, for benchmark points yielding the maximum signal cross section.}
\vspace{5mm}
\begin{tabular}{|c|c|c|c|c|c|c|c|c|c|c|}
\hline
 & \multicolumn{2}{c}{14 TeV} & \multicolumn{2}{|c}{50 TeV} &  \multicolumn{4}{|c|}{100 TeV}  & \multicolumn{2}{|c|}{200 TeV}  \\
 \hline
 & & 3 ab$^{-1}$ & & 30 ab$^{-1}$ & & 3 ab$^{-1}$ & 10 ab$^{-1}$ & 30 ab$^{-1}$ & & 30 ab$^{-1}$ \\
\hline
 & $\sigma_{XS}$ (ab) & $N_{\sigma}$ & $\sigma_{XS}$ (ab) & $N_{\sigma}$ & $\sigma_{XS}$ (ab) & $N_{\sigma}$ & $N_{\sigma}$ & $N_{\sigma}$ & $\sigma_{XS}$ (ab) & $N_{\sigma}$ \\
 \hline
Background & 72,400 & &  505,500 & & 1,323,000 & & & & 3,268,000 & \\
\hline
B1  & 355 & 19.1  & 4,230 & 231 & 12,700 & 131 & 241 & 426  & 34,200 & 658  \\
\hline
B2 & 284 & 17.2 & 3,460 & 211 & 10,500 & 120 & 223 & 376 & 28,400 & 607 \\
\hline
B3 & 93 & 8.8 & 1,260 & 121 & 3,990 & 70.8 & 131 &  226 & 11,100 & 240 \\
\hline
B4 & 59.6 & 7.1 & 865 & 98.7 &  2,790 & 59.4 & 108 & 184 & 7,900 & 313 \\
\hline
B5 & 24.7 & 3.5 & 391 & 56.3 & 1,300 & 35.3 & 63.5 &  110 & 3,800 & 193 \\
\hline
B6 & 13.6 & 2.2 & 233 & 40.1 & 799 & 24.3 & 43.7 & 79.0 & 2,380 & 136 \\
\hline
B7 & 8.9 & 1.6 & 162 & 28.9 & 568 & 19.9 & 36.1 & 64.2 & 1,700 & 109 \\
\hline
B8 & 4.7 & 1.0 & 92.6 & 19.7 & 334 & 13.1 & 24.1 & 41.2 & 1,000 & 73.1 \\
\hline
B9 & 2.0 & 0.45 & 41.6 & 10.4 & 154 & 7.0 & 13.0 &  22.0 & 484 & 39.0 \\
\hline
B10 & 1.0 & 0.27 & 23.1 & 6.0 & 87.2 & 4.3 &  8.1 & 13.4 & 279 & 25.9 \\
\hline
B11 & 0.27 & 0.07 & 6.7 & 1.9 & 26.1 & 1.4 & 2.4 &  4.3 & 85.6 & 9.0 \\
\hline
\end{tabular}
\label{TbbaaMax}
\end{table*} 
 

\end{widetext}

\begin{widetext}
\onecolumngrid

\begin{table*}[t!]
\caption{Cross sections and BDT analysis results for the $b \bar{b} \gamma \gamma$ final state, for benchmark points yielding the minimum signal cross section.}
\vspace{5mm}
\begin{tabular}{|c|c|c|c|c|c|c|c|c|c|c|}
\hline
 & \multicolumn{2}{c}{14 TeV} & \multicolumn{2}{|c}{50 TeV} &  \multicolumn{4}{|c|}{100 TeV}  & \multicolumn{2}{|c|}{200 TeV}  \\
 \hline
 & & 3 ab$^{-1}$ & & 30 ab$^{-1}$ & & 3 ab$^{-1}$ &  10 ab$^{-1}$ &  30 ab$^{-1}$ & & 30 ab$^{-1}$ \\
\hline
 & $\sigma_{XS}$ (ab) & $N_{\sigma}$ & $\sigma_{XS}$ (ab) & $N_{\sigma}$ & $\sigma_{XS}$ (ab) & $N_{\sigma}$ &  $N_{\sigma}$ &  $N_{\sigma}$ & $\sigma_{XS}$ (ab) & $N_{\sigma}$ \\
 \hline
Background & 72,400 & &  505,500 & & 1,323,000  &  &  &  & 3,268,000 & \\
\hline
B1 & 6.3 & 0.50 & 75.2 & 7.3 & 230 & 3.9 & 7.1 &  12.2 & 610 & 19.9  \\
\hline
B2 & 7.8 & 0.66 & 94.8 & 9.2 & 287 & 5.2 & 9.8  & 15.8 & 775 & 24.4 \\
\hline
B3 & 5.8 & 0.70 & 79.4 & 9.9 & 250 & 6.2 & 11.4 & 19.5 & 702 & 32.9 \\
\hline
B4 & 2.4 & 0.36 & 35.8 & 5.7 & 116 & 3.5 & 6.5  & 10.6 & 331 & 18.8 \\
\hline
B5 & 2.2 & 0.41 & 37.2 & 6.9 & 127 & 4.6 & 8.5  & 13.9 & 374 & 25.4 \\
\hline
B6 & 2.8 & 0.52 & 49.1 & 9.6 & 170 & 6.5 & 10.9 & 20.7 & 501 & 34.1 \\
\hline
B7 & 1.7 & 0.32 & 30.9 & 6.2 & 110 & 4.1 & 7.7  & 13.7 & 328 & 23.9 \\
\hline
B8 & 0.97 & 0.20 & 19.4 & 4.3 & 70.5 & 2.7 & 5.0 & 8.6 & 220 & 18.0 \\
\hline
B9 & 0.58 & 0.14 & 12.5  & 3.1 & 46.6 & 2.1 & 3.9 & 6.8 & 147 & 13.2 \\
\hline
B10 & 0.56 & 0.15 & 13.2 & 3.8 & 50.4 & 2.3 & 4.5 & 7.8 & 163 & 15.5 \\
\hline
B11 & 0.21 & 0.06 & 5.2 & 1.5 & 20.4 & 1.1 & 1.8 & 3.1 & 66.9 & 6.9 \\
\hline
\end{tabular}
\label{TbbaaMin}
\end{table*} 
 
\end{widetext}

\section{Appendix B: $4\tau$ Analysis}
\label{AppendixB}

In this section, we display the remaining kinematic distributions used in our BDT analysis which were not included in the main text of section~\ref{4tau}. We also include here tables of cross sections and $N_\sigma$ results from our BDT analysis for the sets of benchmark points yielding the maximum and minimum signal cross section.

\begin{widetext}
\onecolumngrid

\begin{figure*}[h!]
\centering
\mbox{
\subfigure{
\includegraphics[scale=.23]{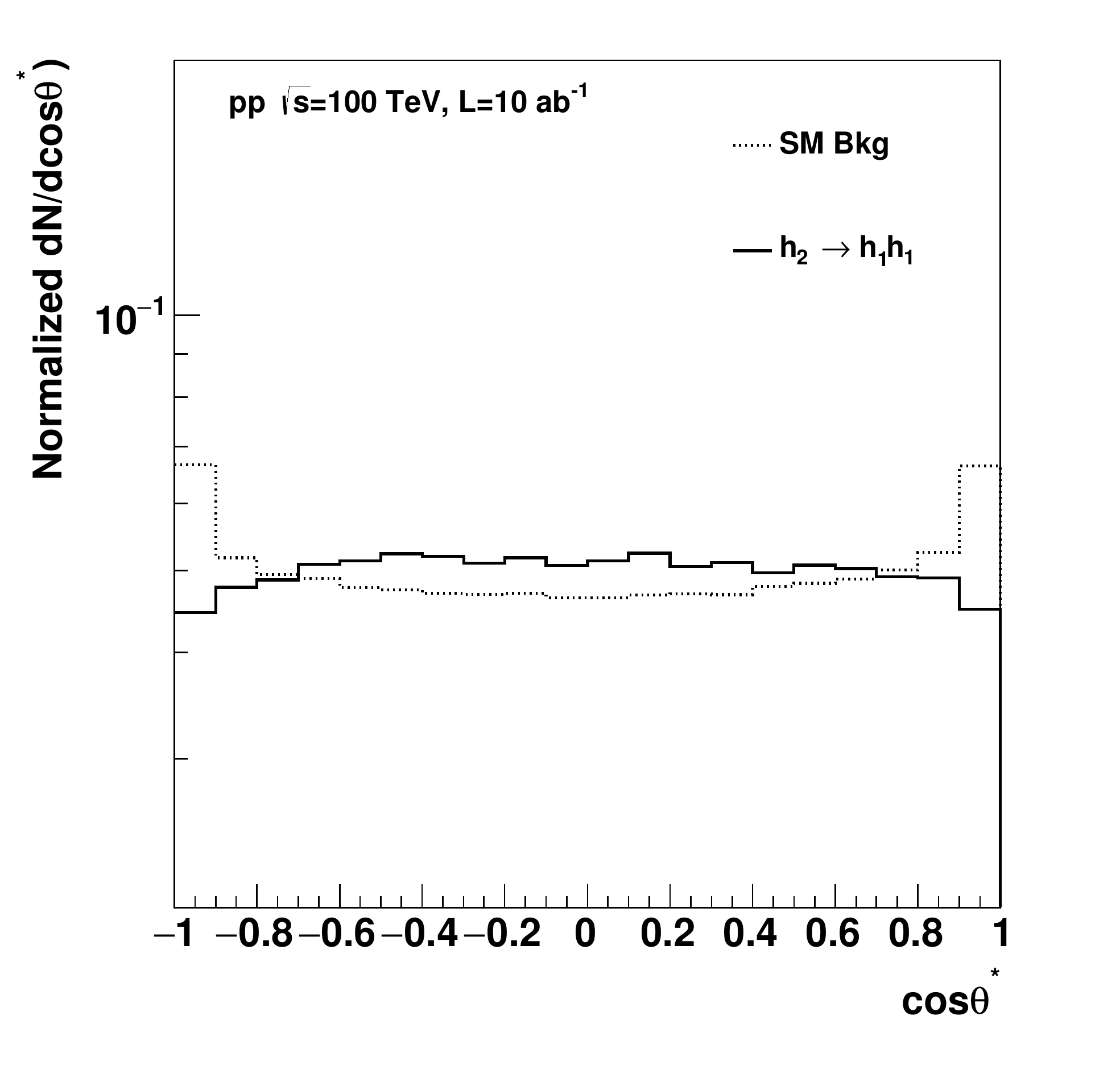}
}
\hspace{-.25in}\subfigure{
\includegraphics[scale=.23]{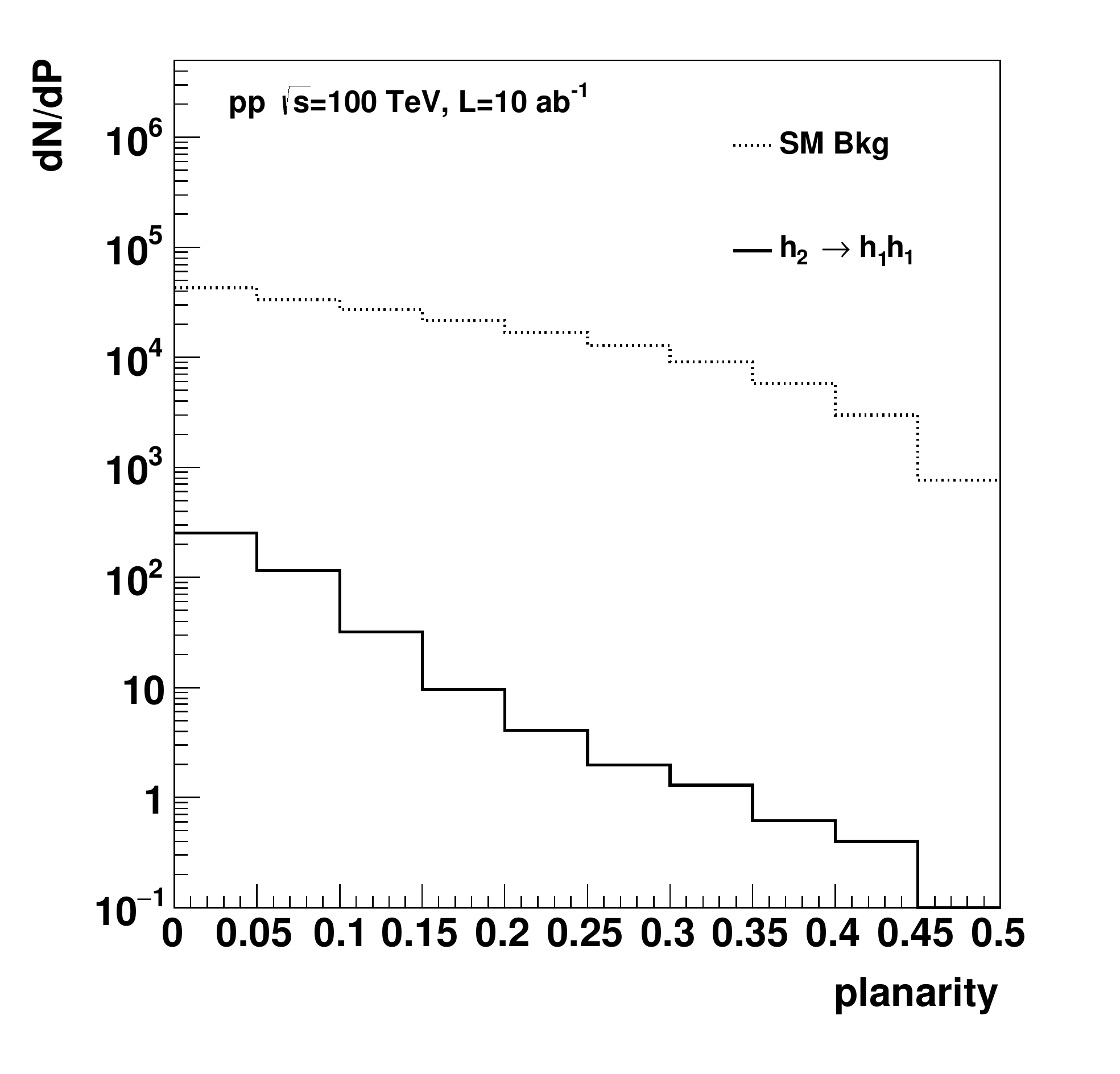}
}
\hspace{-.25in}\subfigure{
\includegraphics[scale=.23]{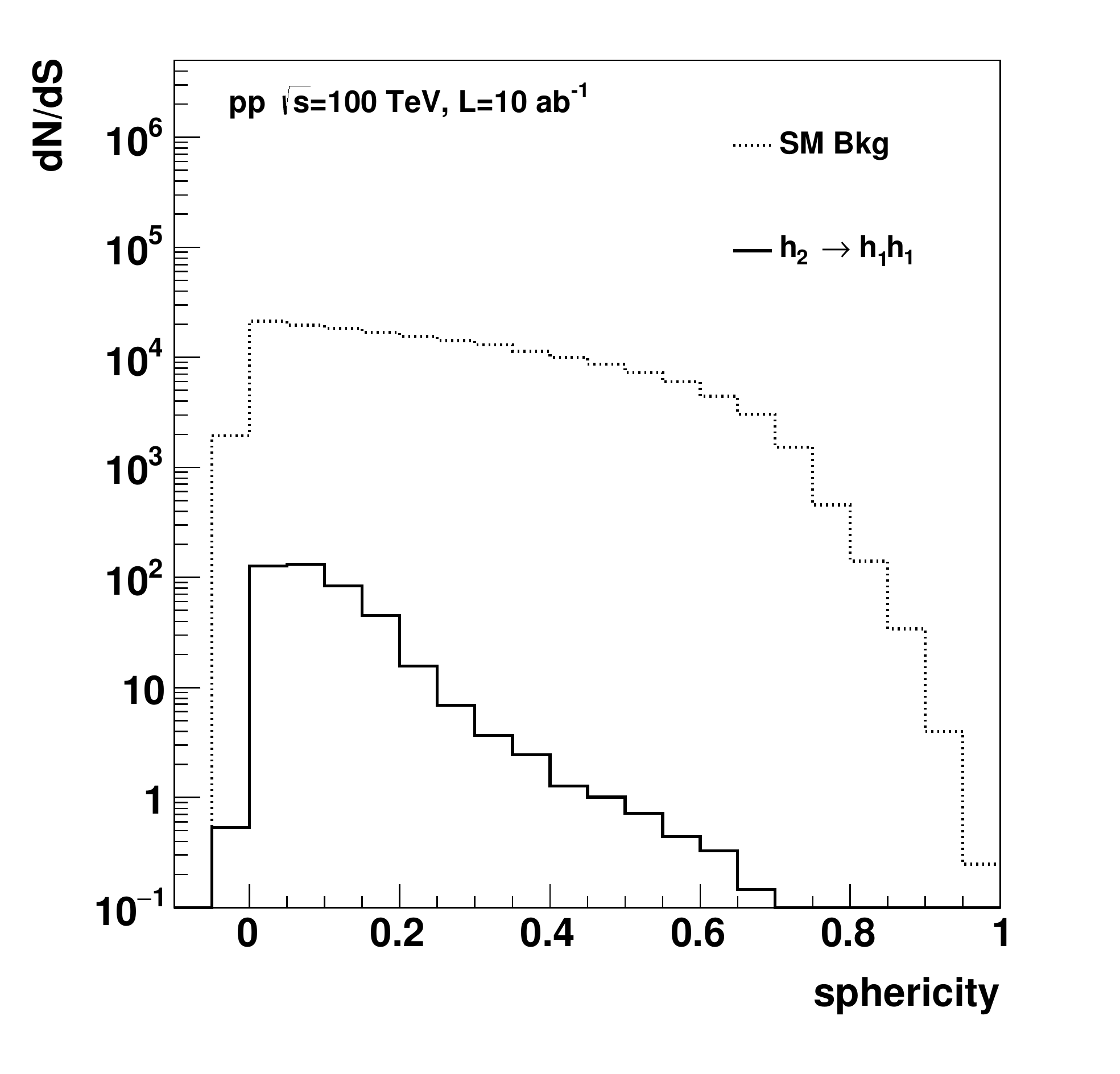}
}
\hspace{-.25in}\subfigure{
\includegraphics[scale=.23]{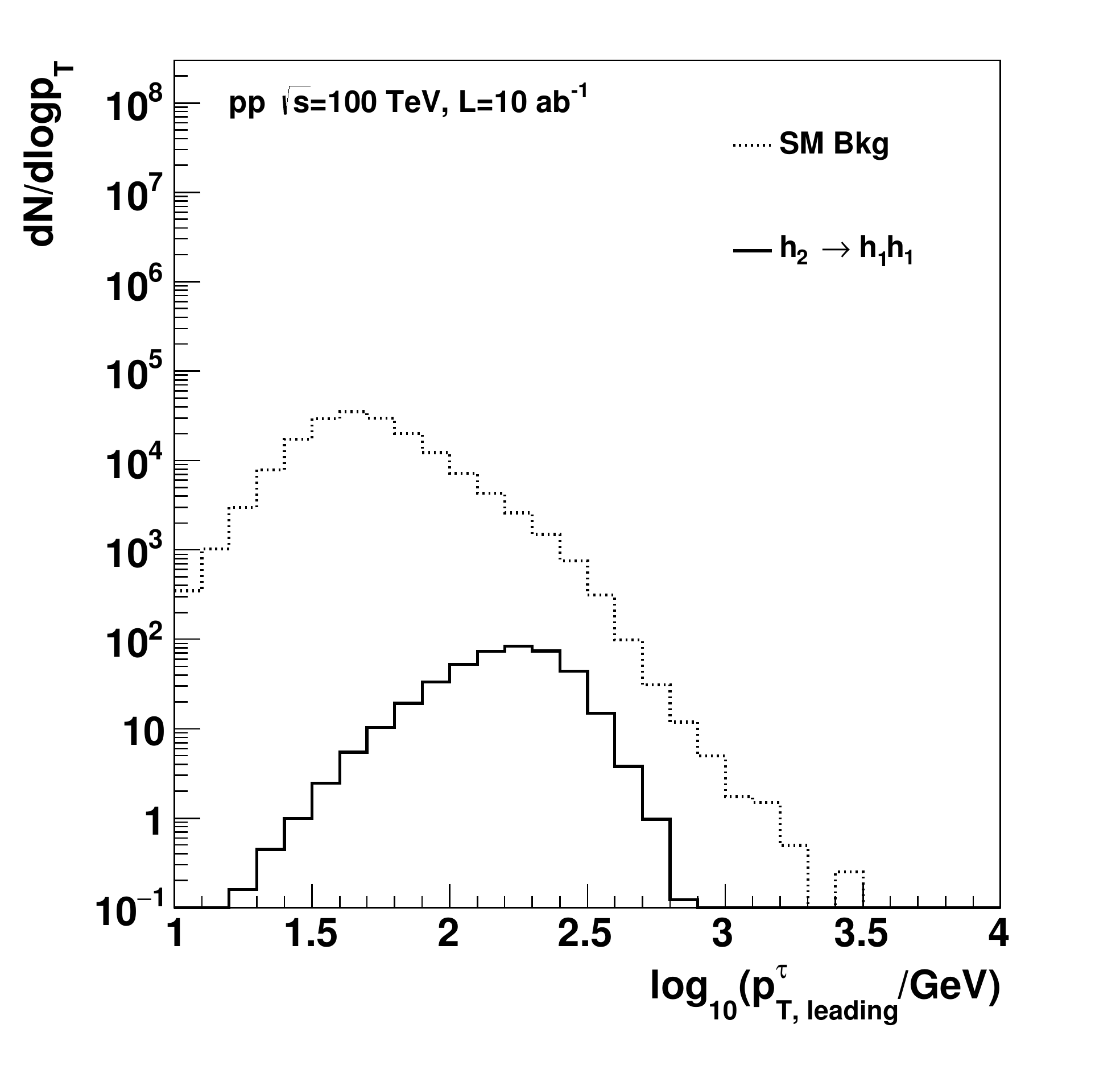}
}
}
\mbox{
\subfigure{
\includegraphics[scale=.23]{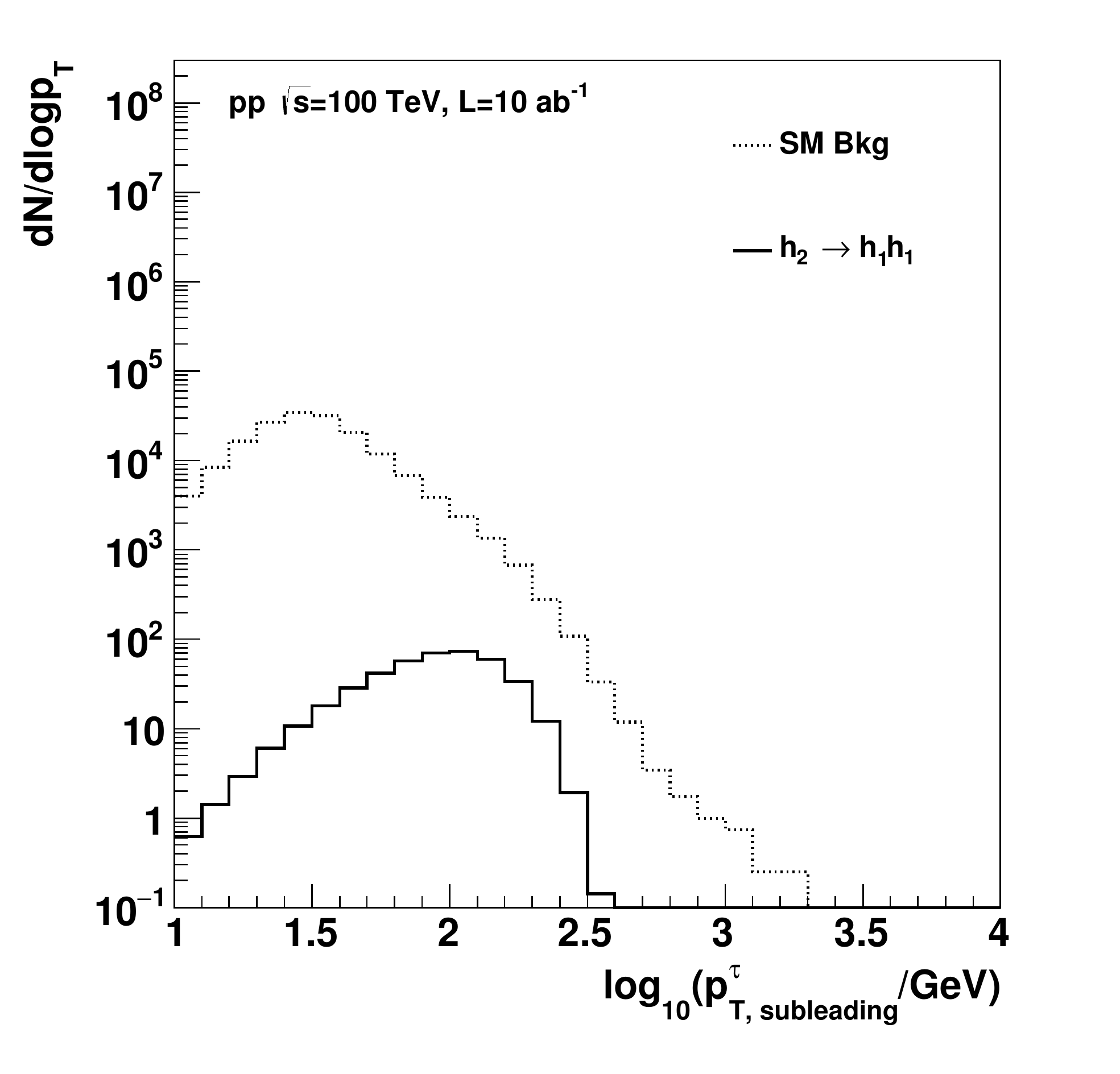}
\hspace{-.25in}\subfigure{
\includegraphics[scale=.23]{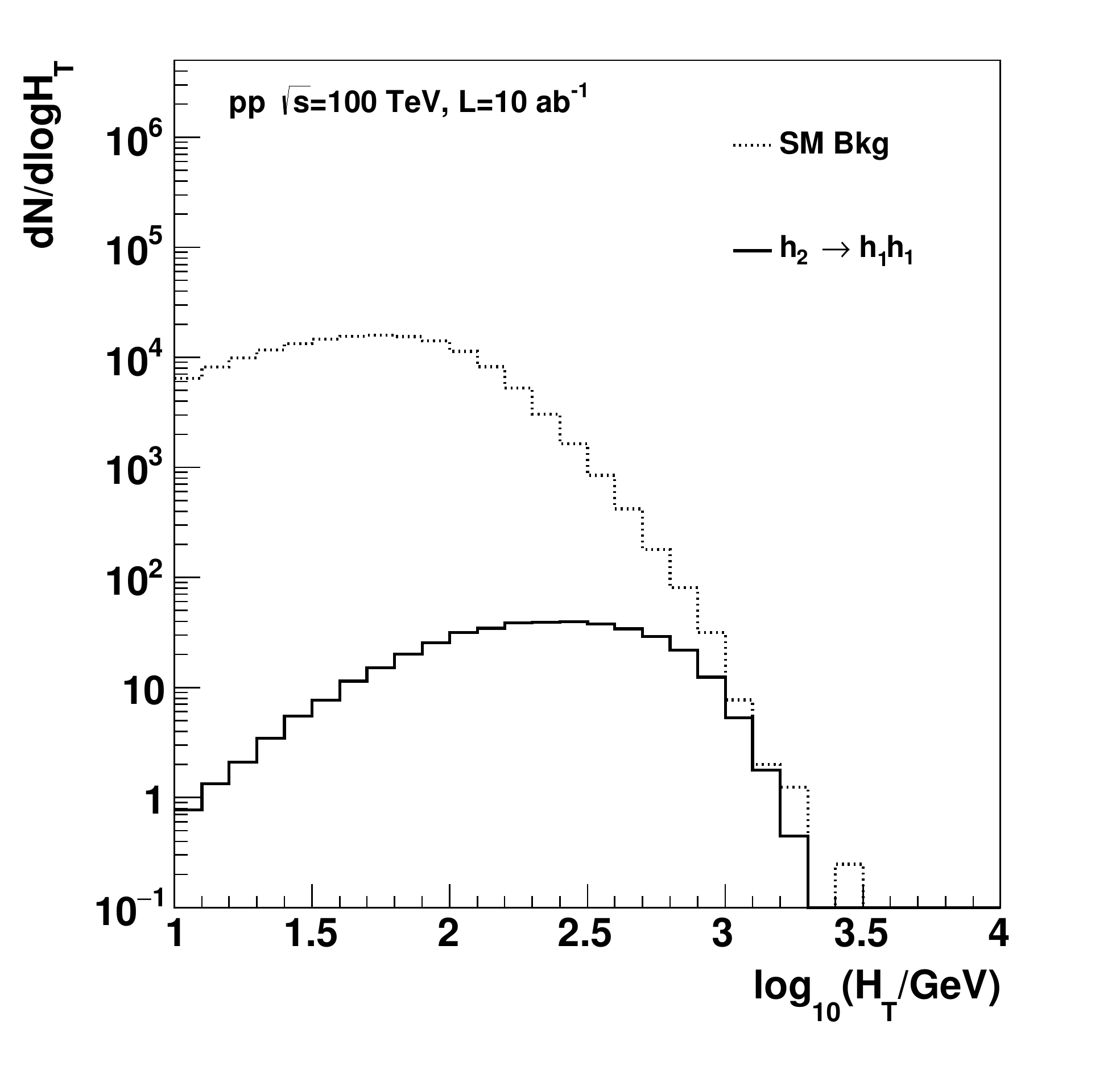}
}
\hspace{-.25in}\subfigure{
\includegraphics[scale=.23]{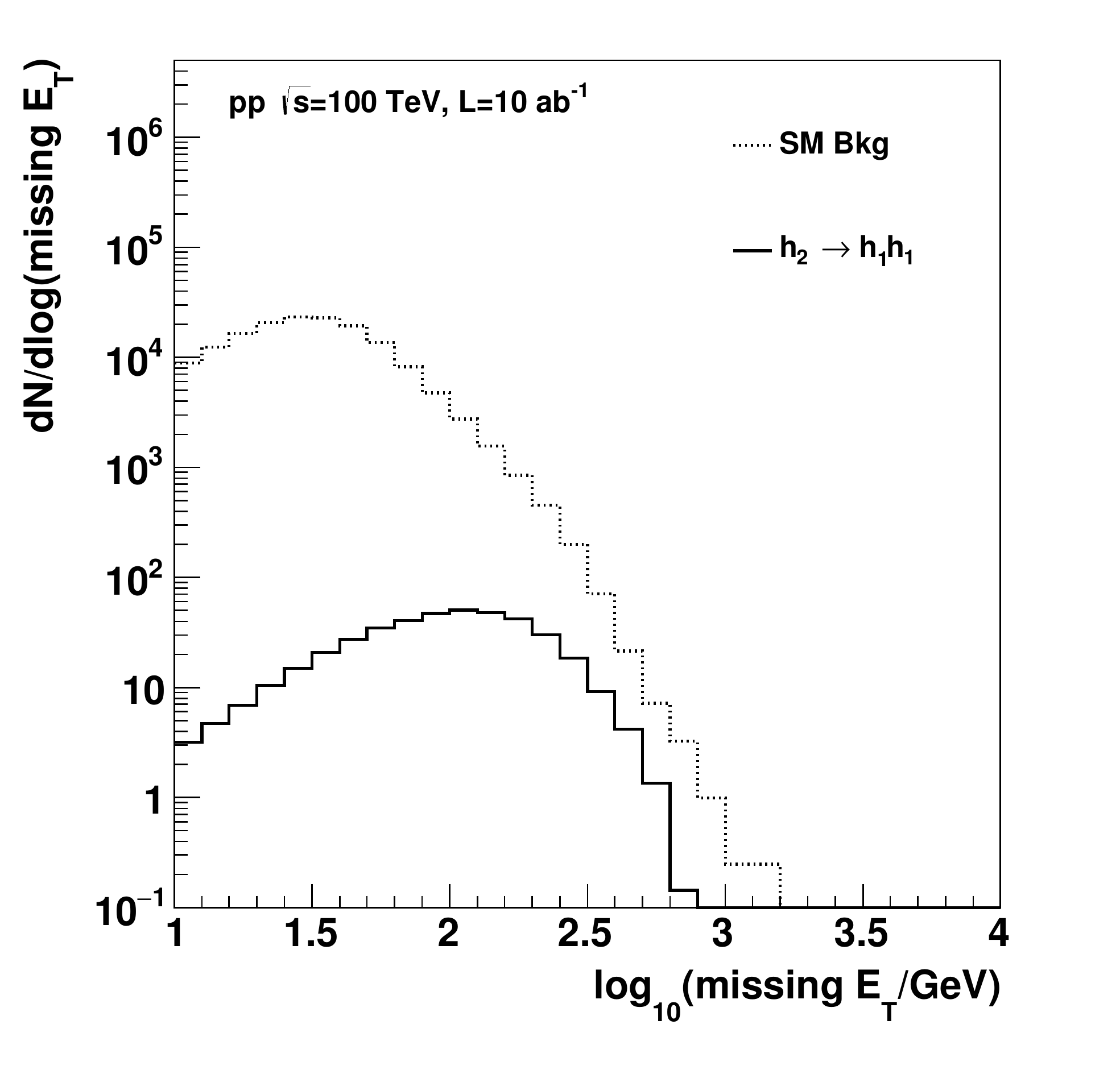}
}
}
}
\caption{Additional kinematics distributions for $4 \tau$ final state, used as inputs to the BDT. The signal distributions correspond  to BM10$^\mathrm{max}$.}
\label{4tauDistrosAdditional}
\end{figure*}

\end{widetext}

\begin{widetext}
\onecolumngrid

\begin{table*}[t!]
\caption{Cross sections and BDT analysis results for the $4 \tau$ final state,  for benchmark points yielding the maximum signal cross section.}
\vspace{5mm}
\begin{tabular}{|c|c|c|c|c|c|c|c|c|c|c|}
\hline
 & \multicolumn{2}{c}{14 TeV} & \multicolumn{2}{|c}{50 TeV} &  \multicolumn{4}{|c|}{100 TeV}  & \multicolumn{2}{|c|}{200 TeV}  \\
 \hline
 & & 3 ab$^{-1}$  &  & 30 ab$^{-1}$ & & 3 ab$^{-1}$ & 10 ab$^{-1}$ & 30 ab$^{-1}$ & & 30 ab$^{-1}$ \\
\hline
 & $\sigma_{XS}$(ab) & $N_{\sigma}$ & $\sigma_{XS}$(ab) & $N_{\sigma}$ & $\sigma_{XS}$(ab)  & $N_{\sigma}$ & $N_{\sigma}$ &  $N_{\sigma}$ & $\sigma_{XS}$(ab) & $N_{\sigma}$ \\
 \hline
Background & 7,500 &  &  30,000 & & 55,000 &  &  &  & 118,000 & \\
\hline
B1  & 457 & 14.1  & 5,440 & 216 & 16,400 & 136 & 248 & 430  & 44,000 & 754  \\
\hline
B2 & 368 & 12.1 & 4,480 & 191 & 13,600 & 121 & 221 & 383 & 36,800 & 668 \\
\hline
B3 & 126 & 5.5 & 1,710 & 96.8 & 5,400 & 64.0 & 117 & 202 & 15,100 & 369 \\
\hline
B4 & 82.9 & 4.1 & 1,200 & 76.2 &  3,890 & 51.8 & 94.6 & 164 & 11,100 & 302 \\
\hline
B5 & 35.2 & 2.1 & 558 & 43.1 & 1,860 & 30.1 & 54.9 & 95.2 & 5,400 & 179 \\
\hline
B6 & 19.7 & 1.3 & 338 & 29.3 & 1,160 & 20.7 & 37.8 & 65.5 & 3,400 & 128 \\
\hline
B7 & 13.1 & 0.92 & 238 & 22.9 &  836 & 16.4 & 30.0 & 52.0 & 2,500 & 100 \\
\hline
B8 & 7.0 & 0.57 & 138 & 14.6 & 497 & 11.0 & 20.2 & 34.9 & 1,540 & 69.0 \\
\hline
B9 & 3.0 & 0.26 & 62.8 & 7.3 & 232 & 5.6 & 10.2 & 17.7 & 731 & 35.6 \\
\hline
B10 & 1.5 & 0.15 & 35.1 & 4.4 & 133 & 3.4 & 6.2 & 10.7 & 426 & 22.9 \\
\hline
B11 & 0.41 & 0.04 & 10.3 & 1.4 & 40.3 & 1.1 & 2.0 & 3.5 & 132 & 7.7 \\
\hline
\end{tabular}
\label{T4tauMax}
\end{table*}

\begin{table*}[t!]
\caption{Cross sections and BDT analysis results for the $4 \tau$ final state, for benchmark points yielding the minimum signal cross section.}
\vspace{5mm}
\begin{tabular}{|c|c|c|c|c|c|c|c|c|c|c|c|}
\hline
 & \multicolumn{2}{c}{14 TeV} & \multicolumn{2}{|c}{50 TeV} &  \multicolumn{4}{|c|}{100 TeV}  & \multicolumn{2}{|c|}{200 TeV}  \\
 \hline
 & & 3 ab$^{-1}$ & & 30 ab$^{-1}$ & & 3 ab$^{-1}$ & 10 ab$^{-1}$ & 30 ab$^{-1}$ & & 30 ab$^{-1}$ \\
\hline
 & $\sigma_{XS}$ (ab) & $N_{\sigma}$ & $\sigma_{XS}$ (ab) & $N_{\sigma}$ & $\sigma_{XS}$ (ab) & $N_{\sigma}$ & $N_{\sigma}$ & $N_{\sigma}$ & $\sigma_{XS}$ (ab) & $N_{\sigma}$ \\
 \hline
Background & 7,500 &  &  30,000 & & 55,000 & &  &   & 118,000 & \\
\hline
B1  & 8.1 & 0.33  & 96.7 & 6.0 & 291 & 4.1 & 7.5 &  12.9  & 784 & 23.5  \\
\hline
B2 & 10.1 & 0.46 & 122 & 7.7 & 371 & 5.3 & 9.5 & 16.5 & 1,003 & 30.4 \\
\hline
B3 & 7.9 & 0.41 & 108 & 7.8 & 341 & 5.5 & 10.0 & 17.4 & 953 & 33.1 \\
\hline
B4 & 3.4 & 0.19 & 49.9 & 4.1 & 162 & 3.0 & 5.4 & 9.3 & 463 & 18.1 \\
\hline
B5 & 3.2 & 0.21 & 53.6 & 5.1 & 183 & 3.7 & 6.8 & 11.8 & 540 & 23.4 \\
\hline
B6 & 4.1 & 0.27 & 71.2 & 6.8 & 245 & 5.1 & 9.3 & 16.1 & 728 & 32.7 \\
\hline
B7 & 2.5 & 0.18 & 45.3 & 4.7 &  159 & 3.6 & 6.5 & 11.3 & 483 & 22.6 \\
\hline
B8 & 1.4 & 0.12 & 29.0 & 3.1 & 105 & 2.4 & 4.4 & 7.7 & 238 & 16.6 \\
\hline
B9 & 0.87 & 0.08 & 18.9 & 2.3 & 70.3 & 1.8 & 3.2 & 5.6 & 223 & 11.7 \\
\hline
B10 & 0.86 & 0.08 & 20.2 & 2.6 & 77.2 & 2.0 & 3.7 & 6.4 & 250 & 14.1 \\
\hline
B11 & 0.32 & 0.03 & 8.1 & 1.1 & 31.5 & 0.90 & 1.6 & 2.8 & 103 & 6.0 \\
\hline
\end{tabular}
\label{T4tauMin}
\end{table*}

\end{widetext}

\begin{table*}[h!]
\begin{tabular}{l|cc}
Decay channel & Branching ratio & Uncertainty \\ \hline \\
$b \bar{b} b \bar{b}$                 &   $3.33\cdot 10^{-1}$   &   $\pm\> 2.20\cdot 10^{-2}$   \\
$\tau \tau b \bar{b}$                 &   $7.29\cdot 10^{-2}$   &   $\pm\> 4.80\cdot 10^{-3}$   \\
$W^{+} (\to l \nu)  W^{-} (\to l \nu) b \bar{b}$   &   $1.09\cdot 10^{-2}$   &   $\pm\> 5.93\cdot 10^{-4}$   \\
$\tau \tau \tau \tau$                 &   $3.99\cdot 10^{-3}$   &   $\pm\> 4.55\cdot 10^{-4}$   \\
$\gamma \gamma b \bar{b}$             &   $2.63\cdot 10^{-3}$   &   $\pm\> 1.58\cdot 10^{-4}$   \\
$W^{+} (\to l \nu)  W^{-} (\to l \nu) \tau \tau$   &   $1.20\cdot 10^{-3}$   &   $\pm\> 8.56\cdot 10^{-5}$   \\
$\gamma \gamma \tau \tau$             &   $2.88\cdot 10^{-4}$   &   $\pm\> 2.19\cdot 10^{-5}$   \\
$b \bar{b} \mu^{+}\mu^{-}$            &   $2.53\cdot 10^{-4}$   &   $\pm\> 1.73\cdot 10^{-5}$   \\
$Z (\to l^{+} l^{-} ) Z(\to l^{+} l^{-}) b \bar{b}$   &   $1.41\cdot 10^{-4}$   &   $\pm\> 7.64\cdot 10^{-6}$   \\
$b \bar{b} Z (\to l^{+} l^{-} ) \gamma$   &   $1.21\cdot 10^{-4}$   &   $\pm\> 1.16\cdot 10^{-5}$   \\
$W^{+} (\to l \nu)  W^{-} (\to l \nu) W^{+} (\to l \nu)  W^{-} (\to l \nu)$   &   $8.99\cdot 10^{-5}$   &   $\pm\> 7.73\cdot 10^{-6}$   \\
$\gamma \gamma W^{+} (\to l \nu)  W^{-} (\to l \nu)$   &   $4.32\cdot 10^{-5}$   &   $\pm\> 2.85\cdot 10^{-6}$   \\
$\tau \tau \mu^{+}\mu^{-}$            &   $2.77\cdot 10^{-5}$   &   $\pm\> 2.29\cdot 10^{-6}$   \\
$Z (\to l^{+} l^{-} ) Z(\to l^{+} l^{-}) \tau \tau$   &   $1.54\cdot 10^{-5}$   &   $\pm\> 1.10\cdot 10^{-6}$   \\
$\tau \tau Z (\to l^{+} l^{-} ) \gamma$   &   $1.32\cdot 10^{-5}$   &   $\pm\> 1.41\cdot 10^{-6}$   \\
$\gamma \gamma \gamma \gamma$         &   $5.20\cdot 10^{-6}$   &   $\pm\> 5.20\cdot 10^{-7}$   \\
$W^{+} (\to l \nu)  W^{-} (\to l \nu) \mu^{+}\mu^{-}$   &   $4.15\cdot 10^{-6}$   &   $\pm\> 3.07\cdot 10^{-7}$   \\
$Z (\to l^{+} l^{-} ) Z(\to l^{+} l^{-}) W^{+} (\to l \nu)  W^{-} (\to l \nu)$   &   $2.31\cdot 10^{-6}$   &   $\pm\> 1.41\cdot 10^{-7}$   \\
$W^{+} (\to l \nu)  W^{-} (\to l \nu) Z (\to l^{+} l^{-} ) \gamma$   &   $1.99\cdot 10^{-6}$   &   $\pm\> 1.98\cdot 10^{-7}$   \\
$\gamma \gamma \mu^{+}\mu^{-}$        &   $9.99\cdot 10^{-7}$   &   $\pm\> 7.80\cdot 10^{-8}$   \\
$\gamma \gamma Z (\to l^{+} l^{-} ) Z(\to l^{+} l^{-})$   &   $5.57\cdot 10^{-7}$   &   $\pm\> 3.67\cdot 10^{-8}$   \\
$\gamma \gamma Z (\to l^{+} l^{-} ) \gamma$   &   $4.78\cdot 10^{-7}$   &   $\pm\> 4.92\cdot 10^{-8}$   \\
$Z (\to l^{+} l^{-} ) Z(\to l^{+} l^{-}) \mu^{+}\mu^{-}$   &   $5.35\cdot 10^{-8}$   &   $\pm\> 3.95\cdot 10^{-9}$   \\
$Z (\to l^{+} l^{-} ) \gamma \mu^{+}\mu^{-}$   &   $4.59\cdot 10^{-8}$   &   $\pm\> 4.96\cdot 10^{-9}$   \\
$Z (\to l^{+} l^{-} ) Z(\to l^{+} l^{-}) Z (\to l^{+} l^{-} ) \gamma$   &   $2.56\cdot 10^{-8}$   &   $\pm\> 2.55\cdot 10^{-9}$   \\
$Z (\to l^{+} l^{-} ) Z(\to l^{+} l^{-}) Z (\to l^{+} l^{-} ) Z(\to l^{+} l^{-})$   &   $1.49\cdot 10^{-8}$   &   $\pm\> 1.28\cdot 10^{-9}$   \\
$Z (\to l^{+} l^{-} ) \gamma Z (\to l^{+} l^{-} ) \gamma$   &   $1.10\cdot 10^{-8}$   &   $\pm\> 1.97\cdot 10^{-9}$   \\
\end{tabular}
\caption{Branching ratios for final states arising from double-Higgs production, with the requirement of leptonic decays of $W$ and $Z$ bosons. }
\label{branchingRatios}
\end{table*}

\bibliographystyle{h-physrev3.bst}
\bibliography{NextGenRefs}

\end{document}